\documentclass[natbib,doc]{apa7}
\usepackage{amsmath,amssymb}
\usepackage{graphicx}
\usepackage{xcolor}
\hypersetup{
    colorlinks,
    linkcolor={red!50!black},
    citecolor={blue!50!black},
    urlcolor={blue!80!black}
}

\title{Topological inference on brain networks across subtypes of post-stroke aphasia} 
\shorttitle{Topological inference on brain networks}

\author{Yuan Wang$^{1*}$, Jian Yin$^2$, Rutvik H. Desai$^3$}
\affiliation{$^1$Department of Epidemiology and Biostatistics, University of South Carolina, U.S.A. \\
                $^2$Department of Biostatistics, Nanjing Medical University, China.\\
                $^3$Department of Psychology, University of South Carolina, U.S.A. \\
                $^*$Correspondence: wang578@mailbox.sc.edu.}

\abstract{Persistent homology (PH) characterizes the shape of brain networks through the persistence features. Group comparison of persistence features from brain networks can be challenging as they are inherently heterogeneous. A recent scale-space representation of persistence diagram (PD) through heat diffusion reparameterizes using the finite number of Fourier coefficients with respect to the Laplace-Beltrami (LB) eigenfunction expansion of the domain, which provides a powerful vectorized algebraic representation for group comparisons of PDs. In this study, we advance a transposition-based permutation test for comparing multiple groups of PDs through the heat-diffusion estimates of the PDs. We evaluate the empirical performance of the spectral transposition test in capturing within- and between-group similarity and dissimilarity with respect to statistical variation of topological noise and hole location. We also illustrate how the method extends naturally into a clustering scheme by subtyping individuals with post-stroke aphasia through the PDs of their resting-state functional brain networks. }

\keywords{Topological data analysis; Persistent homology; Permutation test; Brain network.}

\begin{document} 
\maketitle

\section{Introduction}
\label{sec:intro}

Brain network modeling based on magnetic resonance imaging (MRI) is an effective approach to understand the functions and dysfunctions of the brain. Brain networks have an innate graph structure that have been traditionally studied through graphical or graph theoretic models based on single-scale covariance estimation \citep{Huang2010} or single-scale graph-theoretic measures \citep{Sporns2002,Rubinov2010}. These models effectively characterize brain network topology and have become the norm for brain network analysis. However, it has recently been noted that single-scale models may not be sufficient in capturing the complexity of brain connectivity and multi-scale models are needed \citep{Betzel2017}. On the other hand, a ubiquitous problem in brain network analysis is selection of threshold on edge weights to reveal significant connections within and between brain regions. Arbitrary threshold may cause problem of bias and consistency across studies \citep{Drakesmith2015,Garrison2015}. A multi-scale approach to brain network modeling has thus become widely adapted through persistent homology (PH), which captures multi-scale features of data through invariant topological structures \citep{Edelsbrunner2002}. Using the language of simplicial homology \citep{Hatcher2001}, PH reveals the underlying topological structures of data by their persistence through a dynamic assortment of points, edges, and triangles. The fact that the overall topological changes hold more significance over fleeting structures in PH makes the algorithm particularly robust under the presence of noise and artifacts, thus revealing more insight on network topology than single-scale measures  \citep{Carlsson2009}. Topological characteristics of the dynamic changes through the PH process are summarized through persistence features. 

Current persistence features consist of barcode and persistence diagram (PD), the original descriptors proposed by \citet{Edelsbrunner2002}, and persistence landscape (PL) \citep{Bubenik2015} and persistence image (PI) \citep{Adams2017}, both of which were developed when the demand increased for incorporating persistence features in statistical inference and machine learning models. Persistence features are inherently heterogeneous for noisy samples, even when the samples come from homogeneous underlying data objects. The heterogeneous nature of persistence features means that statistical inference for group comparison is not straightforward. Parametric inference often requires stringent distributional assumptions, which are rarely met by persistence features. So we utilize a nonparametric inference approach. Permutation testing is a standard nonparametric inference procedure for complex data objects and features without clear distributional properties. It is known as the exact test in statistics since the distribution of the test statistic under the null hypothesis can be exactly computed if we can calculate all the test statistics under every possible permutation. It is thus one of the most widely used inference procedures in neuroimaging studies where the data is typically complex in structure and the underlying distributional properties are difficult to quantify \citep{Nichols2002, Simpson2013b, Winkler2016}. However, generating every possible permutation for brain networks with a large number of nodes is still extremely time consuming even for a modest sample size. Standard permutation testing through approximations only reaches a fraction of the exhaustive list of permutations and is computationally intensive. When the total number of permutations is large, various resampling techniques have been proposed to speed up the computation in the past \citep{Nichols2002, Winkler2016}. These resampling methods generate a small fraction of possible permutations and the statistical significance is computed approximately. Neuroimaging studies typically generate 5,000–1,000,000 permutations, less than a fraction of all possible permutations. A few approaches have been developed to overcome the computational bottleneck for permutation testing on persistence features. The {\em exact topological inference} approach allows for fast permutation of monotone functions built on birth or death times in barcodes with respect to the Komogorov-Smirnov (KS) distance \citep{Chung2019a}. This approach has quadratic run time that beats the exponential run time of standard permutation tests and has been extended to compare PLs \citep{Wang2019, Wang2021}. However, the approach is limited to comparing two features and not applicable for comparing between two sets of features. Another rapid permutation test based on transpositions does not require monotonicity and is applicable for comparing two sets of persistence features \citep{Chung2019b, Song2023}. It has allowed us to develop a unified framework for topological inference through heat kernel estimation of PDs.

Inference and learning approaches comparing PDs have been built on confidence band \citep{Fasy2014} to functional representations \citep{Chung2009, Pachauri2011, Bubenik2015, Reininghaus2015, Carriere2015, Chen2015, Adams2017}, as comparing raw PDs consisting of planar scatter points encoding birth and death times of topological structures often require point matching through, for instance, the Hungarian matching algorithm, which quickly becomes formidable for large-scale data. It is also unclear how we may compare two sets of raw PDs. The functional representation approach overcomes the issue of the points on raw PDs having arbitrary locations and provides an effective framework for downstream comparison. In this approach, PDs essentially undergo a smoothing process, in some cases through a scale-space representation from kernels for heat diffusion of Dirac delta functions uniquely representing the points of PD \citep{Reininghaus2015}. However, existing kernel features on PD are typically convoluted, which lacks flexibility when performing resampling-based statistical inference procedures such as permutation testing. A new scale-space representation of PD was recently proposed based on the heat kernel (HK) estimation \citep{Kulkarni2020}, where the upper-triangular domain of PDs is represented using a finite number of Fourier coefficients with respect to the Laplace-Beltrami (LB) eigenfunction expansion of the domain. It provides a powerful vectorized algebraic representation for comparisons of PDs at the same coordinates, foregoing the need for matching across PDs due to their arbitrary point locations. Motivated by a topology-preserving spectral permutation test \citep{Wang2018}, we developed an inference procedure for comparing two sets of PDs estimated by the new scale-space representation by transposing the PD labels \citep{Wang2022}. By updating only the terms in an $L_2$-distance between the mean HK estimates of two sets of PDs involved in each transposition, computation becomes much faster than standard permutation testing that exchanges an arbitrary number of labels in each iteration. This inference procedure generalizes the method developed by \citet{Wang2018} for comparing persistence features of single-trial univariate signals, where the resampling takes place at the signal level and thus cannot be directly applied to images and networks. The inference framework now resamples at the feature level, which allows us to compare PDs of images and networks. We have also extended it to a new topological ANOVA (T-ANOVA) approach to compare across multiple groups of PDs without dimensionality reduction, as well as a topological clustering scheme in application. 

In this study, we establish a topological inference framework through stability of HK estimation on PDs. We evaluate the empirical performance of the spectral permutation test and T-ANOVA in simulation studies in detecting heterogeneous topological noise and hole location across multiple images. We also apply the methods to study topological difference in brain networks across subtypes of individuals with post-stroke aphasia.

\section{Methods}
\label{sec: methods}

\begin{figure}[t!]
  \includegraphics[width = 1\linewidth]{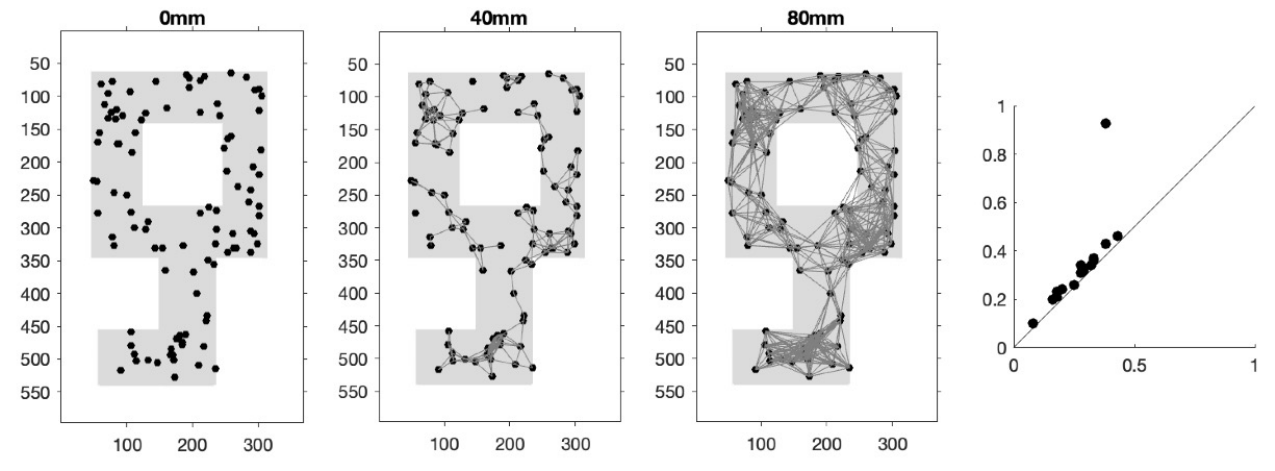}
  \caption{\label{fig: keyhole_pd}Left Three: The evolving 1-skeleton of a 100-point point cloud sampled from a key shape with a hole. Right: PD from the Rips filtration constructed on the 1-skeletons of the point cloud. The point in the PD that corresponds to the key hole stands out with high persistence - much further away from the diagonal ($y = x$) line than the rest of the points.}
\end{figure}

Brain networks are typically modeled as a weighted graph, with the edge weights given by a similarity measure between the measurements on the nodes of the network \citep{Bassett2006,Bien2011}. Suppose we have a network represented by the weighted graph $G = (V, w)$ with the node set $V = \{1,\dots, p\}$ and unique positive undirected edge weights $w=(w_{ij})$ constructed from a similarity measure such as Pearson's correlation. We define the binary network $G_{\epsilon} = (V,w_{\epsilon})$ as a subgraph of $G$ consisting of the node set $V$ and the binary edge weights $w_{\epsilon}$ defined by
\begin{equation}
\label{eq: gf_threshold}
w_{ij,\epsilon} = \left\{ 
\begin{array}{cc}
1 & \text{if}~w_{ij}<\epsilon;\\
0 & \text{otherwise}.
\end{array}\right.
\end{equation} 
As we increase $\epsilon$, which we call the {\em filtration value}, more edges are included in the binary network $G_{\epsilon}$ and so the size of the edge set increases. Since edges connected in the network do not get disconnected again, we observe a sequence of nested subgraphs
\begin{equation}
\label{eq: rips_filtration}
G_{\epsilon_0}\subset G_{\epsilon_1} \subset G_{\epsilon_2} \subset \cdots, 
\end{equation}
for any $$\epsilon_0\le\epsilon_1\le\epsilon_2\le\cdots.$$ This sequence of nested subgraphs make up a {\em Rips filtration} where two nodes with a weight $w_{ij}$ smaller than $\epsilon$ are connected, and the birth and death of {\em homological features} in the form of clusters of nodes and holes formed by more than 3 edges are tracked through the filtration \citep{Lee2011b,Lee2014}. We pair the birth and death times of clusters and holes as the coordinates of scatter points on a planar graph $\{(a_i,b_i)\}_{i=1}^L$ in the {\em persistence diagram} (PD). The persistence of clusters and holes is measured by the drop from their corresponding points to the $y = x$ line on the PD. Long persistence indicates that the corresponding cluster or hole is more likely to be an underlying feature in the network. 
As an illustration in Figure~\ref{fig: keyhole_pd}, we see how a point that corresponds to a hole in a key shape stands out with high persistence in the PD from the Rips filtration constructed on a 100-point point cloud sampled from a key shape with a hole.

\subsection{Heat kernel representation of persistence diagram}
\label{sec: hk}

Since PDs do not form a vector space, they do not possess a natural statistical framework \citep{Bubenik2015} and requires additional manipulation such as kernel smoothing for downstream statistical analysis. As with all noisy data, smoothing is needed for reducing noise (typically random, often artifactual) to better reveal the underlying data structure. We could either smooth data used to construct the networks or smooth persistence  descriptors such as PD. In principle, topological noise and artifacts should be better removed with the latter approach as persistence descriptors are designed to capture topological structures, be they inherent or transient. Another important reason for smoothing PDs is that the heterogenous nature of raw PDs makes it difficult to perform various algebraic operations for statistical inference. Various smoothing methods have been applied to PDs such that statistical inference can be directly performed on them. Beginning with the work of \citet{Chung2009}, each PD is discretized using the the uniform square grid and a concentration map is then obtained by counting the number of points in each pixel, which is equivalent to smoothing PD with a uniform kernel. This approach is somewhat similar to the voxel-based morphometry \citep{Ashburner2000}, where brain tissue density maps are used as a shapeless metric for characterizing concentration of the amount of tissue. \citet{Pachauri2011} followed up the approach by smoothing the PD by a Gaussian kernel centered at every point. Later, \citet{Bubenik2015} proposed the persistence descriptor PL by representing the PD as a function in the Banach space $L_p(\mathbb{R}^2)$ aimed at statistical analysis. PL is easily invertible to a PD, but overemphasizes the high-persistence features. To account for the overall pattern of persistence features, a persistence scale-space (PSS) kernel approach was then proposed by \citet{Reininghaus2015}, where the points in PD are treated as heat sources modeled as Dirac-delta functions and used as an initial condition for a heat diffusion problem with a Dirichlet boundary condition on the diagonal. The closed-form solution of the diffusion problem is an $L_2(\Omega)$ function obtained by convolving the initial condition with a Gaussian kernel, with $\Omega = \{(x,y)\in\mathbb{R}^2: y\ge x\}$ being the closed half plane above the diagonal line $y = x$, and the feature map from the PDs to $L_2(\Omega)$ at a fixed scale yields the PSS kernel. The Hilbert space structure of $L_2(\mathbb{R}^2)$ can be used to construct a PL kernel similar to PSS \citep{Reininghaus2015}. The relatively new persistence descriptor PI sampled at discrete uniform grid to produce homogenous vectorized data out of PDs \citep{Adams2017}. PIs live in Euclidean space and are therefore amenable to a broader range of learning techniques than PLs \citep{Adams2017}. A new heat kernel representation for PDs has recently been proposed by \citet{Kulkarni2020}, which not only allows a non-convoluted vectorized representation for comparisons at the same coordinates of PDs but also smoothing PD at different scales. It has also been combined with transposition test, a novel permutation testing approach, for fast inference on PDs \citep{Wang2022}. We provide in the next two sections a detailed description of heat kernel representation and transposition test on PDs. 

Heat kernel representation has been established as a smoothing framework for noisy measurements on a general manifold $\mathcal{M}\subset\mathbb{R}^d$ \citep{Chung2007, Chung2014.MICCAI}. We assume the fundamental stochastic model
\begin{equation}
\label{eq: model}
f(p) = h(p) + \varepsilon(p), p\in\mathcal{M},
\end{equation}
where $f$ is the noisy measurement, $h$ is the unknown signal, and $\varepsilon$ is a zero-mean Gaussian random field. We make the general enough assumptions that $f\in L_2(\mathcal{M})$, the space of square integrable  functions on $\mathcal{M}$ with the inner product $\langle f_1,f_2\rangle=\int_{\mathcal{M}}f_1(p)f_2(p)d\mu(p)$, where $\mu$ is the Lebesgue measure. A self-adjoint operator $\mathcal{L}$, i.e. $\langle f_1,\mathcal{L}f_2\rangle = \langle \mathcal{L}f_1,f_2\rangle$ for all $f_1, f_2\in L_2(\mathcal{M})$, induces orthonormal eigenvalues $\lambda_k$ and eigenfunctions $\psi_k$ on $\mathcal{M}$: 
\begin{equation}
\label{eq: eigen}
\mathcal{L}\psi_k = \lambda_k\psi_k, k = 0, 1, \dots,
\end{equation}
where, without loss of generality, we can sort the eigenvalues $\lambda_k$ such that $$0 = \lambda_0 \le \lambda_1 \le \cdots,$$ and the eigenfunctions $\psi_k$ can be numerically computed by solving a generalized eigenvalue problem. Then, by Mercer's Theorem, any symmetric positive definite kernel can be written as
\begin{equation}
\label{eq: spd}
K(p,q) = \sum_{k=0}^{\infty}\tau_k\psi_k(p)\psi_k(q)
\end{equation}

Now consider the diffusion-like Cauchy problem
\begin{equation}
\label{eq: diffusion}
\frac{\partial h(\sigma, p)}{\partial \sigma} + \mathcal{L}(\sigma, p) = 0, p\in\mathcal{M},
\end{equation}
with the initial condition $h(\sigma = 0, p) = f(p)$. The partial differential equation \eqref{eq: diffusion} diffuses the noisy data $h(p)$ over $\sigma$. For the self-adjoint operator $\mathcal{L}$, \eqref{eq: diffusion} has the unique solution \citep{Chung2007}
\begin{equation}
\label{eq: diffusion_sol}
h(\sigma,p)=\sum_{k = 0}^{\infty}e^{-\lambda_k\sigma}\langle h,\psi_k\rangle\psi_k(p),
\end{equation}
which provides an estimate $\hat{h}_{\sigma}(p)$ of the unknown signal $h(p)$. The bandwidth $\sigma$ controls the amount of smoothing in the estimate; as $\sigma$ increases, $\hat{h}_{\sigma}(p)$ becomes smoother. When $\mathcal{L}$ is the Laplace-Beltrami (LB) operator, the diffusion equation \eqref{eq: diffusion} becomes the isotropic heat diffusion equation and the kernel \eqref{eq: spd} becomes the heat kernel (HK)
\begin{equation}
\label{eq: hk}
K_{\sigma}(p,q) = \sum_{k=0}^{\infty}e^{-\lambda_k\sigma}\psi_k(p)\psi_k(q), p, q\in\mathcal{M},
\end{equation}
where the $\psi_k$ are the eigenfunctions of the LB operator $\Delta$ satisfying
$$\Delta \psi_k(p) = \lambda_k \psi_k(p)$$
for $p \in \mathcal{M}$. The HK framework has been shown to be equivalent to kernel regression and wavelet \citep{Chung2014.MICCAI}. 

To construct a HK representation of PD, we restrict the domain of diffusion to $\mathcal{M} = \mathcal{T} = \{(x,y)\in\mathbb{R}^2: y>x\}$, i.e. the upper triangular region above the diagonal line $y=x$ where the scatter points of the PD $D = \{(a_i,b_i)\}_{i=1}^P$ are located. We constrain $\mathcal{T}$ 
within a certain range, such as standardizing the coordinates of the PD, so that $\mathcal{T}$ is bounded. Consider heat diffusion equation
\begin{equation}
\label{eq: heat}
\frac{\partial h(\sigma, p)}{\partial \sigma} = \Delta h(\sigma, p)
\end{equation}
with the initial condition
$$\quad h(\sigma=0,p) = \sum_{i=1}^P \delta_{(a_i, b_i)} (p),$$
where $\delta_{(a_i, b_i)}$ is the Dirac-delta function at $(a_i, b_i)$. The scatter points in the PD serve as the heat sources of the diffusion process. To simplify notation, we will refer to any series $h(\sigma,p)$ as $h_{\sigma}(p)$ as the bandwidth $\sigma$ is fixed. A unique solution to \eqref{eq: heat} is given by the HK expansion
\begin{eqnarray}
\label{eq: wfs}
h_{\sigma}(p) &=& \int_{\mathcal{T}} K_{\sigma}(p,q) h_0(q) \;d\mu(q) \nonumber \\
& = &\sum_{k = 0}^{\infty}e^{-\lambda_k\sigma}f_k\psi_k(p),
\end{eqnarray}
where 
\begin{equation}
\label{eq: hk_pd}
K_{\sigma}(p,q) = \sum_{k = 0}^{\infty}e^{-\lambda_k\sigma}\psi_k(p)\psi_k(q), p, q\in\mathcal{T},
\end{equation} 
is the HK with respect to the eigenfunctions $\psi_k$ of the LB operator $\Delta$ satisfying
$\Delta \psi_k(p) = \lambda_k \psi_k(p)$
for $p \in \mathcal{T}$,
and
\begin{equation} 
\label{eq: fourier}
f_k = \big\langle h_0,\psi_k\big\rangle=\int_{\mathcal{T}}  h_0(q)  \psi_k(q) \; d\mu(q)\\
= \sum_{i=1}^P \psi_k(a_i,b_i) 
\end{equation}
are the Fourier coefficients with respect to the the LB eigenfunctions. The first eigenvalue $\lambda_0 =0$ of the LB operator corresponds to eigenfunction $ \psi_0 = \frac{1}{\sqrt{\mu(\mathcal{T})}}$, where $\mu(\mathcal{T})$ is the area of the triangular region $\mathcal{T}$ and $\sigma$ is the bandwidth of the HK.

\begin{figure}[t!]
  \includegraphics[width = 1\linewidth]{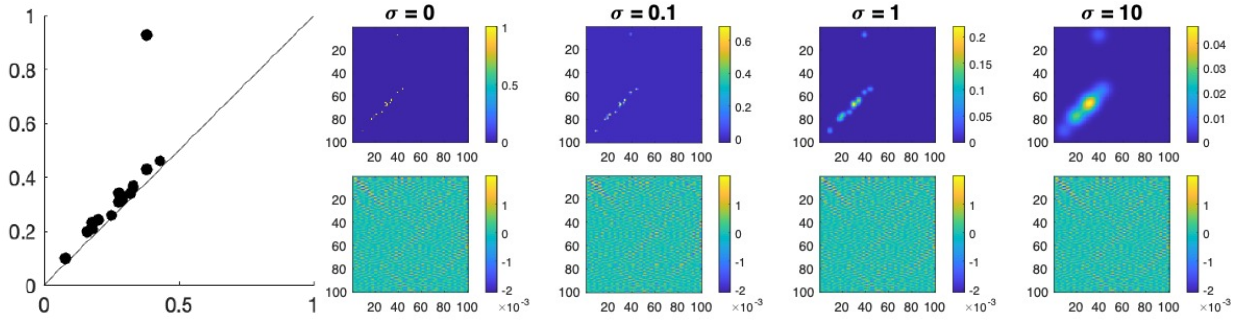}
  \caption{\label{fig: keyhole_pd_bandwidths} Heat kernel (HK) smoothing of the PD,  constructed in Figure~\ref{fig: keyhole_pd}, through Laplace-Beltrami (LB) eigenfunctions with respect to the bandwidths $\sigma = 0$ (original PD), 0.1, 1, 10. Top Row: Smoothed PDs. Bottom Row: Corresponding Fourier coefficients with respect to the LB eigenfunctions presented in matrix form.} 
\end{figure}

The HK expansion \eqref{eq: wfs} provides a vectorized representation of the PD $D$ so that we can compare across PDs at the same coordinates. In practice, we include sufficiently large $\kappa$ terms to approximate the HK expansion:
\begin{equation}
\label{eq: wfs_finite}
h^{\kappa}_\sigma(p) = \sum_{k = 0}^{\kappa}e^{-\lambda_k\sigma}f_k\psi_k(p),
\end{equation}
which we refer to as the degree-$\kappa$ HK estimate of the given PD. When $\sigma = 0$, we can completely recover the initial scatter points. As $\sigma \to \infty$, it is essentially smoothing the PD with a uniform kernel on $\mathcal{T}$. Figure~\ref{fig: keyhole_pd_bandwidths} shows the HK smoothing of a PD with respect to the bandwidths $\sigma = 0, 0.1, 1, 10$. Note that the Fourier coefficients $f_k$ remain the same for all $k$ when constructing the HK expansion at different diffusion scale $\sigma$. 

As a distance measure for the HK-estimated PDs, we use the $L_2$-distance between the functions $h^1, h^2 \in L_2(\mathcal{T})$ defined as
\begin{equation}
\label{eq: l2}
\| h^1 - h^2\|^2_2 = \sum_{k=0}^{\infty}e^{-\lambda_k\sigma}(h^1_{k} - h^2_{k})^2,
\end{equation}  
where the $h^1_{k}$ and $h^2_{k}, k = 0, \dots, \infty,$ are the respective Fourier coefficients of $h^1$ and $h^2$ as defined in \eqref{eq: fourier} with respect to the LB eigenfunctions.

In the standard kernel setup, we have the feature map 
$$\Phi_{\sigma}: \mathcal{D}\rightarrow L_2(\mathcal{T}),$$
where $L_2(\mathcal{T})$ is the space of square integrable functions on $\mathcal{T}$ with the $L_2$-distance 
between the functions $g^1, g^2 \in L_2(\mathcal{T})$ defined as
\begin{equation}
\| g^1 - g^2\|^2_2 = \sum_{k=0}^{\infty}e^{-\lambda_k\sigma}(g^1_{k} - g^2_{k})^2,
\end{equation}  
where the $g^1_{k}$ and $g^2_{k}, k = 0, \dots, \infty,$ are the respective Fourier coefficients of $g^1$ and $g^2$ as defined in \eqref{eq: fourier} with respect to the LB eigenfunctions $\psi_k, k = 0,\dots,\infty$. 
Given bandwidth $\sigma > 0$, 
$$\Phi_{\sigma}(D) = h_{\sigma} = \sum_{k = 0}^{\infty}e^{-\lambda_k\sigma}f_k\psi_k(p), p\in D,$$ 
as defined in \eqref{eq: wfs} for a PD $D\in\mathcal{D}$. This feature map corresponds to the kernel 
$$K_{\sigma}(D_1,D_2)=\langle \Phi_{\sigma}(D_1),\Phi_{\sigma}(D_2)\rangle_{L_2(\mathcal{T})},$$ 
an explicit form of which is given by \eqref{eq: hk_pd}:
\begin{equation}
K_{\sigma}(p,q) = \sum_{k = 0}^{\infty}e^{-\lambda_k\sigma}\psi_k(p)\psi_k(q), p\in D_1, q\in D_2.
\end{equation}
We can show stability of the heat kernel
\begin{equation}
\|K_{\sigma} * g^1 - K_{\sigma} * g^2 \|_2   \leq \|g^1 - g^2\|_2
\end{equation}
as follows: The integral version of Jensen's inequality is
$$\phi \left(\int w(x) \;dx \right) \leq \int \phi ( w(x)) \;dx$$
for convex function $\phi$ \citep{Matkowski1994}. Following Jensen's inequality, 
\begin{eqnarray}
\|K_{\sigma} *g(p)\|_2^2  &=& \int_{\mathcal{T}} \left\vert\int_{\mathcal{T}} K_{\sigma}(p,q) g(q) \; d \mu(q)  \right\vert^2  \;d \mu(p) \\
     &\leq &         \int_{\mathcal{T}} \int_{\mathcal{T}} K_{\sigma}(p,q) \| g(q) \|^2 \; d \mu(q)   \;d \mu(p) \\
     &= & \int_{\mathcal{T}}  \left\vert g(q) \right\vert^2 \int_{\mathcal{T}} K_{\sigma}(p,q)  \; d \mu(p)   \;d \mu(q) \\
     &=&  \int_{\mathcal{T}}  \left\vert g(q) \right\vert^2  \;d \mu(q). 
\end{eqnarray}
We used the fact heat kernel $K_{\sigma}(p,q)$ is a probability distribution such that
$$\int_{\mathcal{T}} K_{\sigma}(p,q)  \; d \mu(p)=1.$$
Hence
$$\| K_{\sigma} *g(p)   \|_2  \leq  \|  g(p) \|_2$$
showing HK smoothing on PD is a contraction map \citep{Chung2018.EMBC}.
Letting $g=g^1-g^2$, we have the stability results. The HK smoothing reduces the topological variability in PD.

\begin{figure}[b!]
  \includegraphics[width = 1\linewidth]{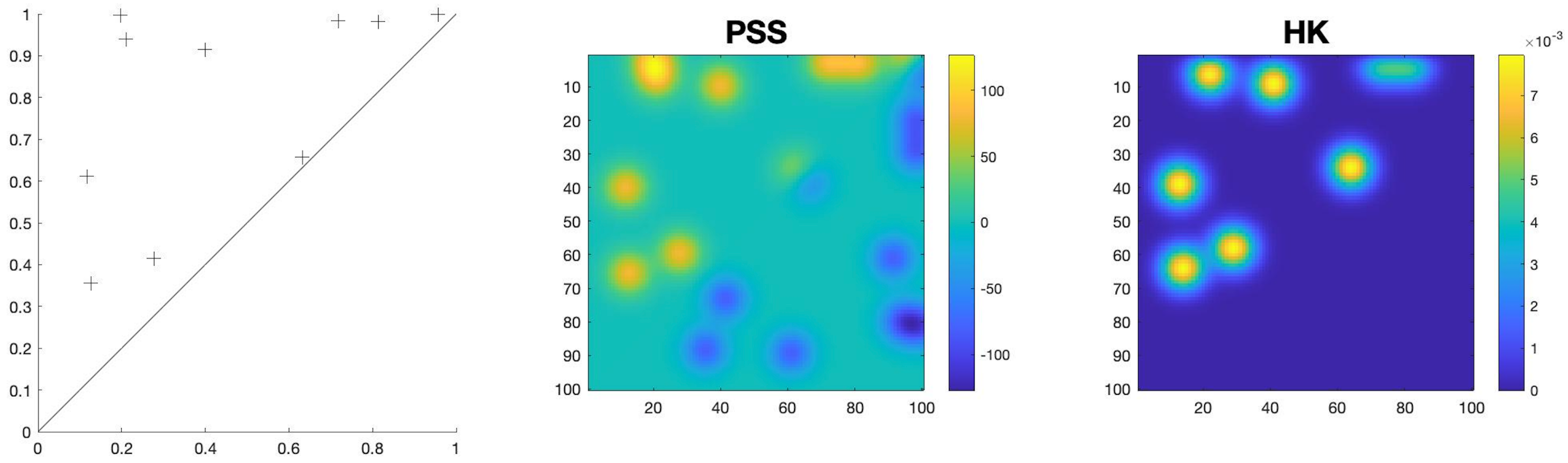}
  \caption{\label{fig: pd_pss_hk}An example of a PD and PSS- and HK-estimated versions.}
\end{figure}

We use a simple example with each of two PDs containing one of the two points $(-\lambda,\lambda)$ and $(-\lambda + 1,\lambda + 1)$ \citep{Reininghaus2015}, as an illustration of the stability of the kernel smoothing procedures. When comparing two PDs, the $L_2$-distance induced by the HK does not weigh over any points in the PDs, as the distance between the two points is $\sum_{k=0}^{\infty}e^{-\lambda_k\sigma}(\psi_k(-\lambda,\lambda)-\psi_k(-\lambda+1,\lambda+1))^2$, which remains constant as $\lambda\rightarrow\infty$. In contrast, the PL-induced kernel distance is dominated by variations in the points of high persistence in the PDs, as the distance between the two points grows in the order of $\sqrt{\lambda}$ and is unbounded, whereas the Wasserstein distance and PSS-induced kernel distance do not over emphasize the high-persistence points, as the distance between the two points asymptotically approach a constant as $\lambda\rightarrow\infty$ \citep{Reininghaus2015}. While the PSS kernel representation, like our HK representation of PD, also uses an exact solution to the heat diffusion problem with the original PD as the initial condition (Figure~\ref{fig: pd_pss_hk}), the implicit form of the solution is difficult to manipulate for cost-effective resampling-based statistical inference. It is likewise difficult to manipulate the Wasserstein distance and PL-induced distance for the same purpose.

\subsection{Permutation test on HK-estimated PDs}
\label{sec: transposition}

Existing kernel features on PD have been shown theoretically and empirically to work well with machine learning frameworks \citep{Reininghaus2015, Adams2017} but are typically convoluted, which lacks flexibility when performing resampling-based statistical inference procedures such as permutation testing. Our past studies have shown powerful applications of the series representation of the heat diffusion problem, such as comparing the persistence features of brain signals through built permutation test based on HK estimates of signals \citep{Wang2018}, where we studied how topology of signals is preserved by permuting Fourier coefficients of sine and cosine basis functions. The approach provides a ground for permutation testing based on spectral components. The downside, however, is the computational load, with spectral permutation of single-trial signals requiring hours on end to converge. 

Here, we use the HK for PD smoothing and subsequent statistical inference based on the HK-estimated PDs. Once we have the HK estimates of PD, we can use them as the basis for statistical inference. Suppose we want to permute the elements of two ordered sets with sizes $m$ and $n$
\begin{eqnarray*}
{\bf x} & = & (x_1,x_2,\dots,x_m), \\
{\bf y} & = & (y_1,y_2,\dots,y_n),
\end{eqnarray*}
in a permutation test with the test statistic $f({\bf x},{\bf y})$. Under the null hypothesis, we assume exchangeability of ${\bf x}$ and ${\bf y}$. Each permutation is an unrestricted rearrangement of the combined ordered set ${\bf z} = (x_1,\dots,x_m,y_1,\dots,y_n)$ and we denote all possible $(m+n)!$ permutations $\mathbb{S}_{m+n}$, which is a symmetric groups of order $m+n$. The standard approximate permutation test typically used in practice is built on on uniform sampling from the full set of permutations. The required number of permutations for convergence increases exponentially as the sample sizes increase. Even with sample sizes like $m=n=20$, the random permutation test requires significant computational resources if we compute the test statistic for each exchange of group labels. 

A {\em transposition} is defined as a permutation $\pi_{ij}$ that exchanges the $i$-th and $j$-th elements between ${\bf x}$ and ${\bf y}$ while keeping all the other elements fixed, i.e.
\begin{eqnarray*}
\pi_{ij}({\bf x}) & = & (x_1,\dots,x_{i-1},y_j,x_{i+1},\dots,x_m),\\
\pi_{ij}({\bf y}) & = & (y_1,\dots,y_{j-1},x_i,y_{j+1},\dots,y_n).
\end{eqnarray*}
Any permutation in $\mathbb{S}_{m+n}$ can be reached by a sequence of transpositions \citep{Chung2019b}. The random transposition is a random walk related to card shuffling problems and it is a special case of walk in symmetric groups \citep{Aldous1983,Aldous1986}. The walk between elements within ${\bf x}$ or ${\bf y}$ is also allowed but will not affect the computation a symmetric test functions. Instead of performing uniform random sampling in $\mathbb{S}_{m+n}$, we can perform a sequence of random walks and compute the test statistic at each walk. Consider walks in the two sample setting. We will determine how test statistic changes over each walk. Over random walk or transposition $\pi_{ij}$, the statistic changes from $L({\bf x}, {\bf y})$ to $L( \pi_{ij} ({\bf x}), \pi_{ij}({\bf y}))$. Instead of computing $L( \pi_{ij} ({\bf x}), \pi_{ij}({\bf y}))$ directly, we can compute it from $L({\bf x}, {\bf y})$ incrementally in {\em constant} run time by updating  the value of $L({\bf x}, {\bf y})$. If $L$ is an algebraic function that only involves addition, subtraction, multiplication, division, integer exponents, there must exists a function $M$ such that 
$$ L(\pi_{ij}({\bf x}), \pi_{ij}({\bf y})) = M (L ({\bf x}, {\bf y}), x_i, y_i), \label{eq:increment}$$
where the computational complexity of $g$ is constant \citep{Chung2019b}. For instance, basic test statistics such as the two-sample $t$-statistic and $F$-statistic are algebraic functions. If we take computation involving  fractional exponents as constant run time, then a much wider class of statistics such as correlations can all have iterative formulation  with constant run time. In the case of computing two-sample $t$-statistic with $m$ and $n$ samples directly, we need to compute the sample means, which takes $O(m)$ and $O(n)$ algebraic operations each. Then need to compute the sample variances and pool them together, which requires $O(3m+2)$ and $O(3n+2)$ operations each. Combining the numerator and denominator in $t$-statistic takes $O(16)$ operations. Thus, it takes total $O(4(m+n)+20)$ operations to compute the $t$-statistic at each permutation. In general, by only updating the terms in the test statistic affected by each transposition, the transposition test would require considerably less computational resources than the standard approximate permutation test. 


When we compare two groups of PDs with sample sizes $m$ and $n$, we assume under the null hypothesis that the functional means of the HK expansion of PDs are the same for both groups, for a fixed bandwidth $\sigma > 0$. The Fourier coefficients in the HK expansion of population PDs in the two groups are unknown. We estimate them with the HK expansion of sample PDs $\{f^{i}\}$ and $\{g^{j}\}$ from the groups approximated by their degree-$\kappa$ estimates: 
\begin{eqnarray}
\label{eq: wfs_pd}
f^i(p) &=& \sum_{k = 0}^{\kappa}e^{-\lambda_{k}\sigma}f^i_{k}\psi_k(p), i = 1,\dots, m,\\
g^j(p) &=& \sum_{k = 0}^{\kappa}e^{-\lambda_{k}\sigma}g^j_{k}\psi_k(p), j = 1,\dots, n,
\end{eqnarray}
where $f^i_{k}$ and $g^j_{k}$, $k = 0, \dots, \kappa$, are the Fourier coefficients with respect to the $k$-th LB eigenfunction $\boldsymbol{\psi}_k$. Their functional means are
\begin{eqnarray}
\label{eq: wfs_mean}
\bar{f}(p) &=& \sum_{k = 0}^{\kappa}e^{-\lambda_{k}\sigma}\bar{f}_{k}\psi_k(p), \\
\bar{g}(p) &=& \sum_{k = 0}^{\kappa}e^{-\lambda_{k}\sigma}\bar{g}_{k}\psi_k(p),
\end{eqnarray}
where $\bar{f}_k = \frac{1}{m}\sum_{i = 1}^m f^i_k$ and $\bar{g}_k = \frac{1}{n}\sum_{j = 1}^n g^j_k$ are the mean Fourier coefficients. We then use the $L_2$-norm difference $\left\|\bar{f} - \bar{g}\right\|^2_2$ between the functional means as a test statistic for measuring the group difference in HK expansion of the PDs. We can algebraically show that
\begin{equation}
\label{eq: wfs_l2}
\left\|\bar{f} - \bar{g}\right\|^2_2 = \sum_{k=0}^{\kappa} e^{-\lambda_k\sigma}(\bar{f}_k-\bar{g}_k)^2.
\end{equation}

In a standard approximate permutation test, the subject labels of the two groups are randomly exchanged. Here, we build the permutation test on transposition $\pi_{ij}$ that only exchanges the $i$-th and $j$-th subject labels between $\{f^i, i = 1,\dots, m\}$ and $\{g^j, j = 1,\dots,n\}$ and keeps all the other PDs fixed, i.e.
\begin{eqnarray}
\label{eq: trans}
\pi_{ij}(f^{1},\dots, f^{m}) & = & (f^{1},\dots, g^{j},\dots,f^{m}), \\
\pi_{ij}(g^{1},\dots, g^{n}) & = & (g^{1},\dots, f^{i}, \dots, g^{n}),
\end{eqnarray}
which we call a {\em spectral transposition}. 
Any permutation of the two groups of $m$ and $n$ subjects is reachable by a sequence of transpositions, which has been shown to be computationally much more efficient than the standard permutation testing procedure of exchanging all labels at once \citep{Chung2019b}. We generate the empirical distribution for the permutation test through the spetral transpositions. In one spectral transposition $\pi_{ij}$, we obtain the $L_2$-distance between the functional means of the degree-$\kappa$ HK estimates of PDs based on transposed labels:
\begin{equation}
\label{eq: l2_trans}
L_2(f,g) = \left\|\bar{f}' - \bar{g}'\right\|^2_2 = \sum_{k=0}^{\kappa} e^{-\lambda_k\sigma}(\bar{f}'_k-\bar{g}'_k)^2,
\end{equation}
where $$\bar{f}'_k=\bar{f}_k+\frac{1}{m}(g^j_k-f^i_k) \mbox{ and }  \bar{g}'_k=\bar{g}_k+\frac{1}{n}(f^i_k-g^j_k)$$ are the means of transposed Fourier coefficients. Since we know $\bar{f}_k$ and $\bar{g}_k$ already, we simply update the terms $\frac{1}{m}(g^j_k-f^i_k)$ and $\frac{1}{n}(f^i_k-g^j_k)$ affected by the transposition. 
The $p$-value of the spectral permutation test is then calculated as the proportion of $L_2$-distances in the empirical distribution exceeding the $L_2$-distance between the observed PDs. To ensure convergence, we perform upward of 100,000 permutations until the $p$-value stabilizes. 

\subsection{Topological analysis of variance via transpositions on HK-estimated PDs}

Topological analysis of variance allows us to assess within- and between-group similarity and dissimilarity in PDs across multiple groups. The challenge of applying an ANOVA procedure to raw PDs is that they do not have unique means \citep{Mileyko2011}. Thus, \citet{Heo2012} applied the standard ANOVA procedure to raw PDs reduced in dimensionality via Isomap. In contrast, our HK-estimates of PDs have well-defined functional means and $L_2$-distance through Fourier coefficients, which provides a natural framework for topological analysis of variance on PDs without any dimensionality reduction beforehand. 

To describe our heuristics in constructing an effective topological ANOVA framework, suppose the $K$ groups of HK-estimated PDs are expressed as follows:  
$$
\begin{array}{ccccc}
\text{Group 1}: & f^{11} & f^{12} & \cdots & f^{1n_1} \\
\text{Group 2}: & f^{21} & f^{22} & \cdots & f^{2n_2} \\
\vdots & \vdots \\
\text{Group K}: & f^{K1} & f^{K2} & \cdots & f^{Kn_K}
\end{array}
$$
Motivated by the standard ANOVA procedure, we could try and build an $F$-statistic 
comparing $K$ groups of HK-estimated PDs through the $L_2$-distance in \eqref{eq: wfs_l2}. A topological between-group sum of squares could take the form of 
\begin{equation}
\label{eq: tssb}
\sum_{i = 1}^{K}n_i||\bar{f}^i-\bar{f}||^2_2,
\end{equation}
and a topological within-group sum of squares the form of 
\begin{equation}
\label{eq: tssw}
\sum_{i = 1}^{K}\sum_{j = 1}^{n_i}||f^{ij} - \bar{f}^i||^2_2,
\end{equation}
where $f^{ij}$ is the HK-estimate of the $j$-th PD of the $i$-th group, $\bar{f}^i$ is the functional mean of the HK-estimates of PDs in the $i$-th group, and $\bar{f}$ is the grand functional mean over the HK-estimates of all PDs. The functional means would serve as the topological centroids. Ideally the $F$-statistic would follow $F$-distribution under some mild normality assumptions on the HK-estimated PDs, 
such as
$$
\frac{\sum_{i = 1}^{K}n_i||\bar{f}^i-\bar{f}||^2_2/K-1}{\sum_{i = 1}^{K}\sum_{j = 1}^{n_i}||f^{ij} - \bar{f}^i||^2_2/N-K}\sim F_{K-1,N-K},
$$
with $N = \sum_{i = 1}^{K}n_i$ and 
$$\sum_{i = 1}^{K}n_i||\bar{f}^i-\bar{f}||^2_2 \sim \chi^2_{K-1},$$ 
$$\sum_{i = 1}^{K}\sum_{j = 1}^{n_i}||f^{ij} - \bar{f}^i||^2_2 \sim \chi^2_{N-K}.$$
However, normality assumptions for heterogeneous features like PDs may be too strong to satisfy on multivariate data. 

Instead of fiddling with parametric constraints, we use a permutational ANOVA approach that bypasses the distributional issue and has found significant applications on multivariate data in response to complex experimental designs of ecological studies, where variables usually consist of counts of counts, percentage cover, frequencies, or biomass for a large number of species, and many other fields including chemistry, social sciences, agriculture, medicine, genetics, psychology, economics  \citep{Anderson2001, Anderson2017}. 
Here we build our test statistic for the permutational ANOVA based on pre-calculated pairwise distances between PDs so that no recalculation of distances is required after each transposition. We will only need to update the within- and between-group sums of distances after each transposition. We will refer to our topological ANOVA procedure as {\bf T-ANOVA}, where we define the topological between-group sum of squares (TSSB) and topological within-group sum of squares (TSSW) based on sums of pairwise $L_2$-distances: 
\begin{eqnarray}
\label{eq: tss}
\text{TSSB} & = & \sum_{\substack{i,i'=1\\i<i'}}^{K}\sum_{j,j'} ||f^{ij}-f^{i'j'}||^2_2 \\
\text{TSSW} & = & \sum_{i=1}^K\sum_{j<j'}||f^{ij}-f^{ij'}||^2_2. 
\end{eqnarray}
We measure the between- and within-group disparity with the ratio statistic
\begin{equation}
\label{eq: ratio}
\phi = \frac{\text{TSSB}}{\text{TSSW}}.
\end{equation}
In each transposition, we randomly sample the group labels $i_1$ and $i_2$ out of the $K$ groups with respect to the proportions of the group sizes $n_i/N$. We then uniformly sample the subject labels $j_1$ and $j_2$ out of the $i_1$-th and $i_2$-th group respectively for transposition. We can prove by induction that any permutation between the groups can be reached by a sequence of transpositions through Theorem 1 in \citep{Chung2019b} showing any permutation between two groups can be reached by a sequence of transpositions.

In a transposition, we only update the pairwise $L_2$-distances in TSSB and TSSW affected by the transposition: 
\begin{eqnarray}
\text{TSSW}' & = & \sum_{i=1}^K\sum_{j<j'}||f^{ij}-f^{ij'}||^2_2\nonumber\\
& + & \sum_{j'\ne j_2}||f^{i_1j_1}-f^{i_2j'}||^2_2  - \sum_{j'\ne j_1}||f^{i_1j_1}-f^{i_1j'}||^2_2\label{eq: tssw_trans1}\\
& + & \sum_{j'\ne j_1}||f^{i_2j_2}-f^{i_1j'}||^2_2 - \sum_{j'\ne j_2}||f^{i_2j_2}-f^{i_2j'}||^2_2\label{eq: tssw_trans2}\\
& = & \text{TSSW}\nonumber\\
& + & \sum_{j'\ne j_2}||f^{i_1j_1}-f^{i_2j'}||^2_2  - \sum_{j'\ne j_1}||f^{i_1j_1}-f^{i_1j'}||^2_2\nonumber\\
& + & \sum_{j'\ne j_1}||f^{i_2j_2}-f^{i_1j'}||^2_2 - \sum_{j'\ne j_2}||f^{i_2j_2}-f^{i_2j'}||^2_2\nonumber,
\end{eqnarray}
where we adjust terms involving only groups $i_1$ and $i_2$ with \eqref{eq: tssw_trans1} and \eqref{eq: tssw_trans2}.
\begin{eqnarray}
\text{TSSB}' & = & \sum_{\substack{i,i'=1\\i<i'}}^{K}\sum_{j,j'} ||f^{ij}-f^{i'j'}||^2_2 \nonumber\\
& - & \sum_{j'\ne j_2}||f^{i_1j_1}-f^{i_2j'}||^2_2  + \sum_{j'\ne j_1}||f^{i_1j_1}-f^{i_1j'}||^2_2\label{eq: tssb_trans1}\\
& - & \sum_{j'\ne j_1}||f^{i_2j_2}-f^{i_1j'}||^2_2 + \sum_{j'\ne j_2}||f^{i_2j_2}-f^{i_2j'}||^2_2\label{eq: tssb_trans2}\\
& + & \mathbb{1}(i'\ne i_1, i_2)\sum_{j'=1}^{n_{i'}}(||f^{i_2j_2}-f^{i'j'}||^2_2 - ||f^{i_1j_1}-f^{i'j'}||^2_2)\label{eq: tssb_trans3}\\
& + & \mathbb{1}(i'\ne i_1, i_2)\sum_{j'=1}^{n_{i'}}(||f^{i_1j_1}-f^{i'j'}||^2_2 - ||f^{i_2j_2}-f^{i'j'}||^2_2)\label{eq: tssb_trans4}\\
& = & \text{TSSB}\nonumber\\
& - & \sum_{j'\ne j_2}||f^{i_1j_1}-f^{i_2j'}||^2_2  + \sum_{j'\ne j_1}||f^{i_1j_1}-f^{i_1j'}||^2_2\nonumber\\
& - & \sum_{j'\ne j_1}||f^{i_2j_2}-f^{i_1j'}||^2_2 + \sum_{j'\ne j_2}||f^{i_2j_2}-f^{i_2j'}||^2_2\nonumber,
\end{eqnarray}
where we adjust terms involving only groups $i_1$ and $i_2$ with \eqref{eq: tssb_trans1} and \eqref{eq: tssb_trans2}, terms involving groups other than $i_2$ that are affected by $i_1$ with \eqref{eq: tssb_trans3}, and terms involving groups other than $i_1$ that are affected by $i_2$ with \eqref{eq: tssb_trans4}. The ratio statistic is then updated to
\begin{equation}
\label{eq: ratio_trans}
\phi' = \frac{\text{TSSB}'}{\text{TSSW}'}.
\end{equation}

The $p$-value of the T-ANOVA test is then calculated as the proportion of $\phi'$ in the empirical distribution exceeding the $\phi$ between the observed PDs. We keep the transposed labels as the current labels on which we build the next transposition and randomize all labels every 500 transpositions to improve convergence rate.


\section{Performance Evaluation} 
\label{sec: simulations}

We conduct two sets of simulation studies to evaluate performance of the two-sample transposition test and T-ANOVA.

\subsection{Performance of two-sample transposition test}

We investigate how the spectral transposition test detects underlying topological similarity and dissimilarity at the presence of topological noise and artifact. 

\begin{figure}[t!]
  \includegraphics[width = 1\linewidth]{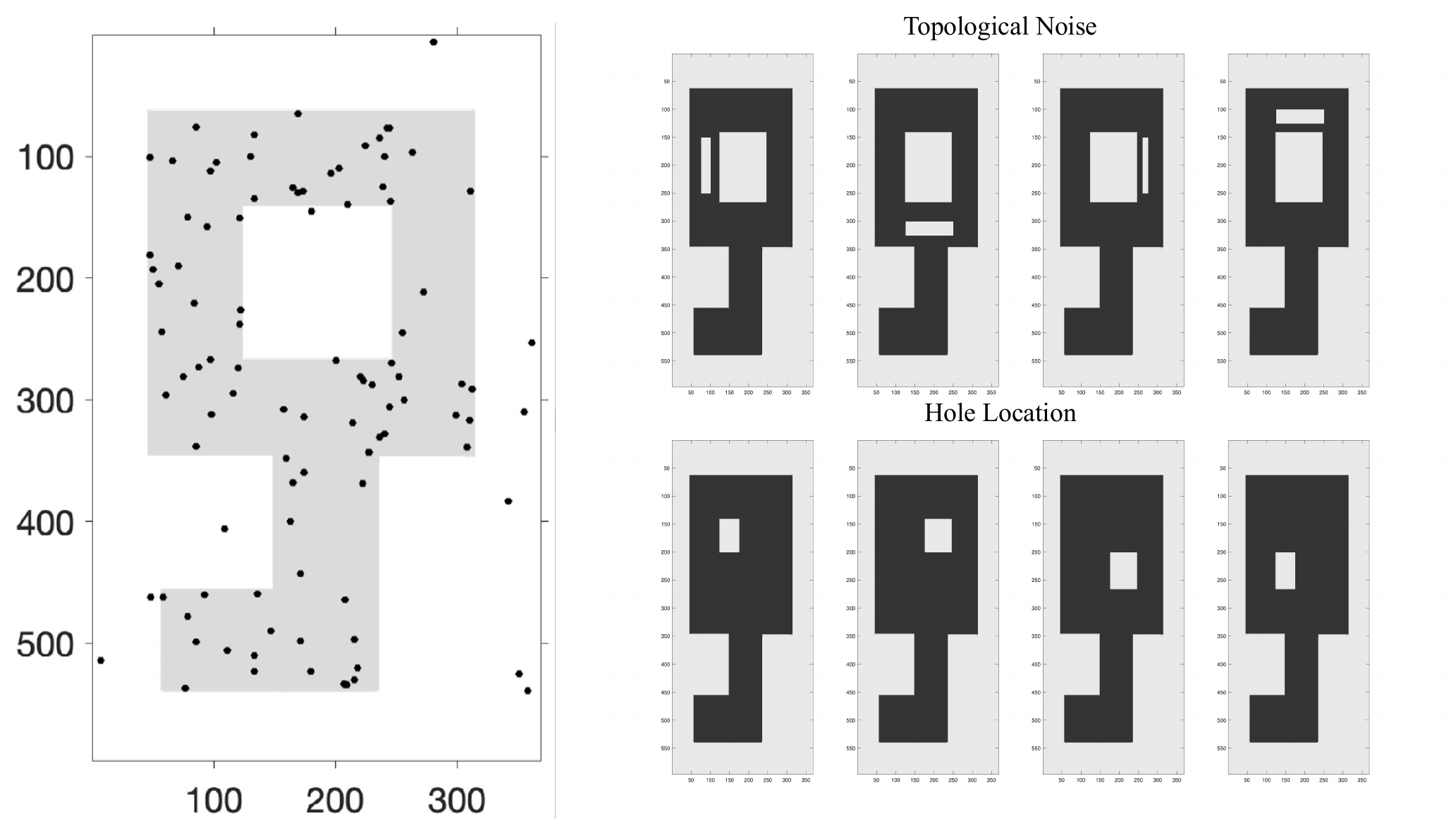}
  \caption{\label{fig: sim_key}Left: We randomly sample 100 points from the image with an innate shape of a key. Top Right Row: Underlying key shape and possible locations of topological noise in the form of a small hole. Bottom Right Row: Variants of the key shape in Group 2. They could appear in the 4 pre-specified forms or randomly out of the variants. }
\end{figure}

\subsubsection{Power of detecting hole in structure} 

We evaluate the power of the transposition test in detecting a key shape with a distinct hole (Figure~\ref{fig: sim_key} left), under different heterogeneity conditions. In each simulation, two groups of five 100-point point clouds are generated: the 100 points in each point cloud of the first group are generated randomly from the part of the rectangular image, whereas the 100 points in each point cloud of the second group are generated randomly with a varied percentage (90\%, 95\%, 100\%) of points from the shape of the key. Rips filtration is constructed on each point cloud. The proposed spectral permutation test is then applied to compare the PDs of the Rips filtrations in the two groups. When there are respectively 90\%, 95\%, and 100\% points sampled from the shape of the key in the second group, the spectral permutation test rejects ($p$-value $<$ 0.05) the null hypothesis of no group difference in 91, 100, and 100\% of 100 simulations (corresponding means $\pm$ standard deviations of $p$-values: 0.0124$\pm$0.0327, 0.0041$\pm$0.0125, 0.0008$\pm$0.0057, showing that the test stays sensitive in detecting the group shape difference when points in the second group are not entirely sampled from the shape of the key.

\subsubsection{Robustness of performance under variation of topological noise and hole location} \label{sec: sim_trans_robust} We conduct two studies to assess the robustness of the test when the underlying topological structure is 'contaminated' with heterogeneous topological noise and when the underlying structure undergoes non-topological changes. 

We first evaluate the robustness of performance under heterogeneity of topological noise. In each of 100 simulations, we use the spectral transposition test to compare Group 1 of $m$ random samples with a varied percentage (90\%, 95\%, 100\%) of 100 points from the original key shape with Group 2 of $n$ random samples from the key shape 'contaminated' with topological noise in the form of a much smaller hole next to the keyhole with pre-specified (in such case $m = 4$ vs $n = 4$) or random locations (in such case $m = 5$ vs $n = 5$, $m = 20$ vs $n = 20$, or $m = 100$ vs $n = 100$). Figure~\ref{fig: sim_key} (top right row) shows the 4 possible locations of the topological noise in Group 2. We expect the test to stay robust to this topological noise. Table~\ref{tab: results_trans_robustness} summarizes the results for different percentage of points when the topological noise appears at pre-specified vs random locations. The spectral transposition test stays robust to the topological noise in fixed and random locations.

  

  \begin{table}[t!]
  \begin{tabular}{ccc}
  \hline
  \multicolumn{3}{c}{{\bf Robustness under Variation of Topological Noise}}\\\hline
  Percentage  & Pre-specified Loc. ($m = 4$ vs $n = 4$) & Random Loc. ($m = 5$ vs $n = 5$) \\\hline
  100\% & 0.4567$\pm$0.2874 & 0.4133$\pm$0.2482\\
  95\% & 0.4777$\pm$0.2843 & 0.4498$\pm$0.2844\\
  90\% & 0.4455$\pm$0.2791 & 0.5214$\pm$0.2983\\\hline
    Percentage  & Random Loc. ($m = 20$ vs $n = 20$) & Random Loc. ($m = 100$ vs $n = 100$) \\\hline
  100\% & 0.5060$\pm$0.3163 & 0.4328$\pm$0.2764\\
  95\% & 0.5016$\pm$0.2998 & 0.4193$\pm$0.2863\\
  90\% & 0.4827$\pm$0.2919 & 0.5260$\pm$0.2812\\\hline
  \multicolumn{3}{c}{{\bf Robustness under Variation of Hole Location}} \\\hline
    Percentage & Pre-specified Loc. ($m = 4$ vs $n = 4$) & Random Loc. ($m = 5$ vs $n = 5$) \\\hline
  100\% & 0.2917$\pm$0.2624 & 0.5005$\pm$0.2883\\
  95\% & 0.2973$\pm$0.2407 & 0.5342$\pm$0.2775\\
  90\% & 0.3065$\pm$0.2505 & 0.4434$\pm$0.3050\\\hline
  Percentage & Radom Loc. ($m = 20$ vs $n = 20$) & Random Loc. ($m = 100$ vs $n = 100$) \\\hline
  100\% & 0.4998$\pm$0.2901 & 0.3845$\pm$0.2620\\
  95\% & 0.4608$\pm$0.2999 & 0.3924$\pm$0.2777\\
  90\% & 0.4810$\pm$0.2568 & 0.4550$\pm$0.2835\\\hline
  \end{tabular}
  \caption{\label{tab: results_trans_robustness}Summary of mean$\pm$standard deviation of $p$-values from the spectral transposition test in 100 simulations. Top half: In each simulation, the test is used to compare a group of $m$ random samples with a varied percentage (90\%, 95\%, 100\%) of 100 points from the original key shape with a group of $n$ random samples with the same percentage of 100 points from the key shape with topological noise in the form of a much smaller hole next to the keyhole. The location of the smaller hole in each random sample of the second group can be pre-specified or randomly chosen from the pre-specified options. Bottom half: In each simulation, the test is used to compare a group of $m$ random samples with a varied percentage (90\%, 95\%, 100\%) of 100 points from the original key shape with only the top left quarter of the keyhole left, with a group of $n$ random samples with the same percentage of 100 points from the original key shape with a random quarter of the keyhole left.}
  \end{table}
  
We then evaluate the robustness of performance under variation of hole location. In each of 100 simulations, we use the spectral transposition test to compare Group 1 of $m$ random samples with a varied percentage (90\%, 95\%, 100\%) of 100 points from the original key shape with only the top left quarter of the keyhole left, with Group 2 of $n$ random samples with the same percentage of 100 points from the original key shape with a pre-specified (in such case $m = 4$ vs $n = 4$) or random (in such case $m = 5$ vs $n = 5$, $m = 20$ vs $n = 20$, or $m = 100$ vs $n = 100$) quarter of the keyhole left. Figure~\ref{fig: sim_key} (bottom right half) shows the 4 possible variants of the keyhole in Group 2. We expect the test to stay robust to this change in structure, which is not topological in nature. Table~\ref{tab: results_trans_robustness} summarizes the results for different percentage of points when the variants appears at pre-specified vs random locations. The spectral transposition test stays robust to the structural variants in fixed and random locations. 


 \subsubsection{Computational time} The computational time of the spectral transposition test grows steadily as the group sample sizes grow. The mean time for each simulation run for $m = 4, 5$ vs $n = 4,5$ between 7 and 10 seconds and standard deviation within 3 seconds. For $m = 20$ vs $n = 20$, the mean time for each simulation run is between 8 and 10 seconds and standard deviation within 3 seconds. For $m = 100$ vs $n = 100$, the mean time for each simulation run is between 9 and 11 seconds and standard deviation within 3 seconds.

\subsection{Performance of T-ANOVA}

In each of the simulation studies in this section, we test the performance of the T-ANOVA in comparing three groups of point clouds simulated under different settings. The performance is compared against the standard PERMANOVA test \citep{Anderson2001}, as well as the topological analysis of variance test proposed by \citet{Heo2012} that runs the univariate ANOVA on dimensionality-reduced PDs by Isomap. 

\begin{figure}[t!]
  \includegraphics[width = 1\linewidth]{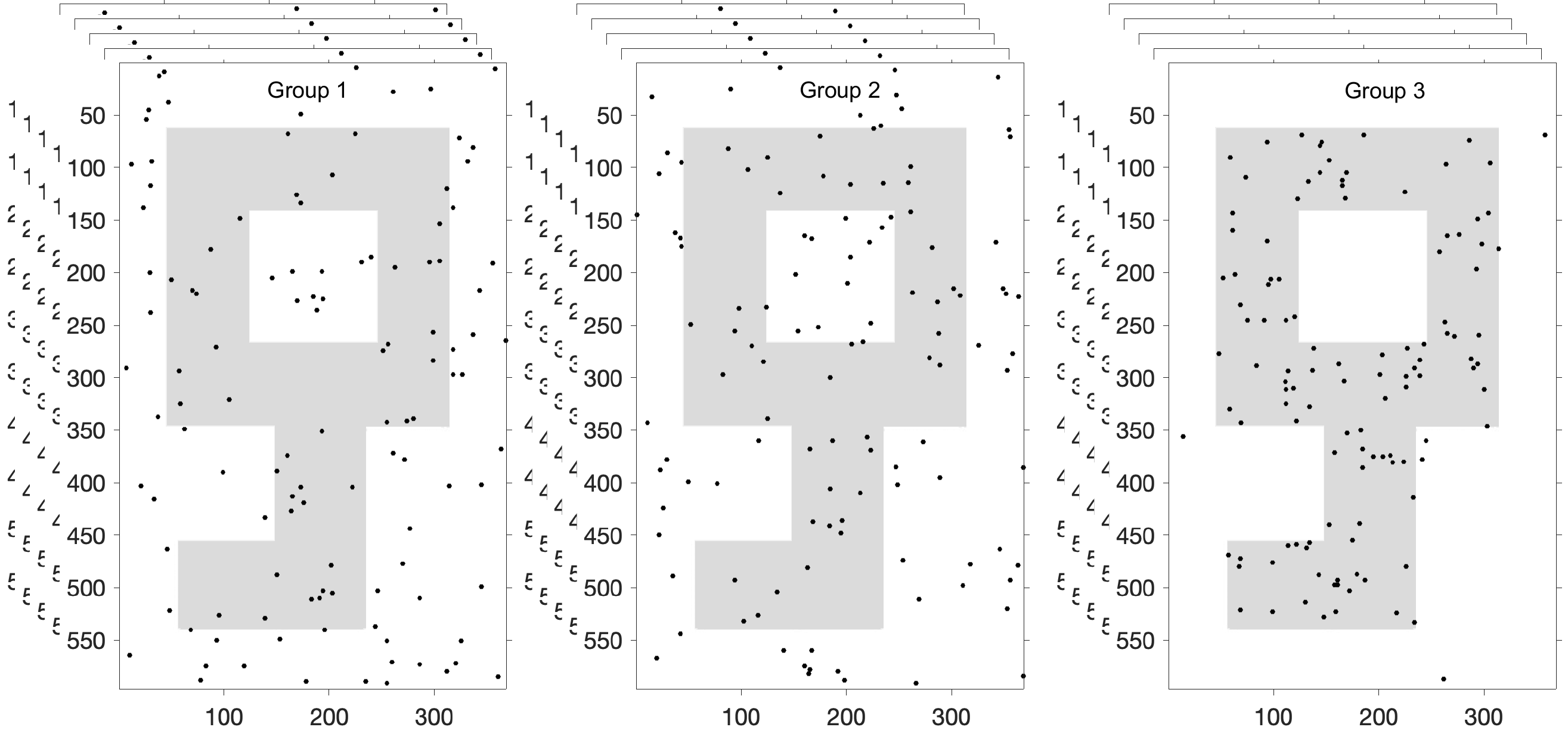}
  \caption{\label{fig: sim_anova_power}An example of $n_1=n_2=n_3=5$ 100-point point clouds where the 100 points in each point cloud of the first two groups are generated randomly from the part of the rectangular image, whereas the 100 points in each point cloud of the third group are generated randomly with 95\% of points from the shape of the key.}
\end{figure}

\subsubsection{Sensitivity in detecting differential hole presence among multiple groups} In each simulation, three groups of $n_1, n_2, n_3$ 100-point point clouds are generated, where the 100 points in each point cloud of the first two groups are generated randomly from the part of the rectangular image, 
whereas the 100 points in each point cloud of the third group are generated randomly with a varied percentage (90\%, 95\%, 100\%) of points from the shape of the key (Figure~\ref{fig: sim_anova_power}). Table~\ref{tab: results_anova_power} shows the results of the T-ANOVA test  in comparison with the other tests.

\begin{table}[b!]
  \begin{tabular}{cccc}
  \hline
   \multicolumn{4}{c}{{\bf Sensitivity in Detecting Differential Hole Presence Among Multiple Groups}} \\\hline
   \multicolumn{4}{c}{$n_1 = n_2 = n_3 = 5$} \\\hline
    Percentage of Points in Key Shape & T-ANOVA & Heo's ANOVA & PERMANOVA  \\\hline
  100\% & $0.0009 \pm 0.0016$ & $0.0038 \pm 0.0101$    & $0.0025 \pm 0.0033$\\
  95\% & $0.0016 \pm 0.0033$ & $0.0109 \pm 0.0258$     & $0.0045 \pm 0.0064$\\
  90\% & $0.0060 \pm 0.0146$ & $0.0436 \pm 0.0791$     & $0.0192 \pm 0.0438$\\ \hline
    \multicolumn{4}{c}{$n_1 = n_2 = n_3 = 20$}\\\hline
    Percentage of Points in Key Shape & T-ANOVA & Heo's ANOVA & PERMANOVA  \\\hline
  100\% & $0.0000 \pm 0.0000$ & $0.0000 \pm 0.0000$	& $0.0000 \pm 0.0000$\\
  95\% & $0.0000 \pm 0.0000$ & $0.0000 \pm 0.0000$	& $0.0000 \pm 0.0000$\\
  90\% & $0.0000 \pm 0.0000$ & $0.0000 \pm 0.0001$	& $0.0000 \pm 0.0000$\\ \hline
    \multicolumn{4}{c}{$n_1 = 5, n_2 = 20, n_3 = 100$} \\\hline
    Percentage of Points in Key Shape & T-ANOVA & Heo's ANOVA & PERMANOVA  \\\hline
  100\% & $0.0253 \pm 0.0699$ & $0.0353 \pm 0.0101$   & $0.0011 \pm 0.0055$\\
  95\% & $0.0485 \pm 0.0827$ 	& $0.0511 \pm 0.1255$   & $0.0028 \pm 0.0117$\\
  90\% & $0.0998 \pm 0.1377$ 	& $0.1281 \pm 0.2395$   & $0.0091 \pm 0.0238$\\ \hline
  \end{tabular}
  \caption{\label{tab: results_anova_power} Summary of mean$\pm$standard deviation of $p$-values 
  from the T-ANOVA, Heo's ANOVA, and PERMANOVA in 100 simulations. In each simulation, three groups of $n_1, n_2, n_3$ 100-point point clouds are generated, where the 100 points in each point cloud of the first two groups are generated randomly from the part of the rectangular image, whereas the 100 points in each point cloud of the third group are generated randomly with a varied percentage (90\%, 95\%, 100\%) of points from the shape of the key.}
  \end{table}

\begin{table}[t!]
  \begin{tabular}{cccc}
  \hline
     \multicolumn{4}{c}{{\bf Robustness under Variation of Topological Noise}} \\\hline
  \multicolumn{4}{c}{Pre-specified Location: $n_1 = n_2 = n_3 = 4$}\\\hline
   Percentage of Points in Key Shape & T-ANOVA & Heo's ANOVA & PERMANOVA  \\\hline
  100\% & $0.4860 \pm 0.2607$ & $0.5267 \pm 0.2675$   & $0.4901 \pm 0.2965$\\
  95\% & $0.4897 \pm 0.2608$ 	& $0.5211 \pm 0.2822$   & $0.5050 \pm 0.2851$\\
  90\% & $0.4974 \pm 0.2963$  & $0.5163 \pm 0.3125$   & $0.4619 \pm 0.2667$\\ \hline
  \multicolumn{4}{c}{Random Location: $n_1 = n_2 = n_3 = 5$} \\\hline
     Percentage of Points in Key Shape & T-ANOVA & Heo's ANOVA & PERMANOVA  \\\hline
  100\% & $ 0.5275 \pm 0.3052$ & $0.4988 \pm 0.2882$   & $0.5518 \pm 0.2952$\\
  95\% & $0.5166 \pm 0.3138$  & $0.5062 \pm 0.2830$    & $0.5308 \pm 0.2870$\\
  90\% & $0.4970 \pm 0.2975$	& $0.5205 \pm 0.2689$    & $0.4923 \pm 0.2928$\\ \hline
  \multicolumn{4}{c}{Random Location: $n_1 = n_2 = n_3 = 20$}\\\hline
     Percentage of Points in Key Shape & T-ANOVA & Heo's ANOVA & PERMANOVA \\\hline
  100\% & $0.4915 \pm 0.2624$	& $0.4812 \pm 0.2640$   & $0.4727 \pm 0.2720$\\
  95\% & $0.4860 \pm 0.2862$  & $0.4864 \pm 0.3072$   & $0.5141 \pm 0.2734$\\
  90\% & $0.4349 \pm 0.2759$  & $0.5139 \pm 0.2791$   & $0.4706 \pm 0.2996$\\ \hline
  \multicolumn{4}{c}{Random Location: $n_1 = 5, n_2 = 20, n_3 = 100$} \\\hline
     Percentage of Points in Key Shape & T-ANOVA & Heo's ANOVA & PERMANOVA  \\\hline
  100\% & $0.5374 \pm 0.2937$ & $0.5001 \pm 0.3028$   & $0.4593 \pm 0.2836$\\
  95\% & $0.5413 \pm 0.2747$  & $0.5016 \pm 0.2738$   & $0.4692 \pm 0.3037$\\
  90\% & $0.4938 \pm 0.2855$ 	& $0.5014 \pm 0.2955$   & $0.5085 \pm 0.3040$\\ \hline
\multicolumn{4}{c}{{\bf Robustness under Variation of Hole Location}} \\\hline
  \multicolumn{4}{c}{Pre-specified Location: $n_1 = n_2 = n_3 = 4$}\\\hline
   Percentage of Points in Key Shape & T-ANOVA & Heo's ANOVA & PERMANOVA \\\hline
  100\% & $0.4887 \pm 0.2873$   & $0.4664 \pm 0.2537$   & $0.5057 \pm 0.2862$\\
  95\% & $0.4948 \pm 0.2776$ 	  & $0.4479 \pm 0.2889$   & $0.5109 \pm 0.2779$\\
  90\% & $0.4364 \pm 0.2572$	  & $0.4651 \pm 0.2916$   & $0.4463 \pm 0.2767$\\ \hline
  \multicolumn{4}{c}{Random Location: $n_1 = n_2 = n_3 = 5$} \\\hline
   Percentage of Points in Key Shape & T-ANOVA & Heo's ANOVA & PERMANOVA \\\hline
  100\% & $0.4785 \pm 0.2737$	 & $0.5505 \pm 0.2706$   & $0.4932 \pm 0.3036$\\
  95\% & $0.5039 \pm 0.3082$ 	 & $0.4872 \pm 0.3055$   & $0.4572 \pm 0.3079$\\
  90\% & $0.4887 \pm 0.2919$   & $0.5302 \pm 0.2784$   & $0.4095 \pm 0.2784$\\ \hline
  \multicolumn{4}{c}{Random Location: $n_1 = n_2 = n_3 = 20$}\\\hline
   Percentage of Points in Key Shape & T-ANOVA & Heo's ANOVA & PERMANOVA\\\hline
  100\% & $0.5249 \pm 0.3118$ & $0.4839 \pm 0.2928$   & $0.5048 \pm 0.3159$\\
  95\% & $0.5292 \pm 0.3082$  & $0.4528 \pm 0.2790$   & $0.5398 \pm 0.3044$\\
  90\% & $0.5183 \pm 0.3009$  & $0.5419 \pm 0.2801$   & $0.5198 \pm 0.2870$\\ \hline
  \multicolumn{4}{c}{Random Location: $n_1 = 5, n_2 = 20, n_3 = 100$} \\\hline
   Percentage of Points in Key Shape & T-ANOVA & Heo's ANOVA & PERMANOVA \\\hline
  100\% & $0.4411 \pm 0.2541$ & $0.4738 \pm 0.3197$   & $0.5481 \pm 0.2707$\\
  95\% & $0.4498 \pm 0.2774$ 	& $0.4623 \pm 0.2825$   & $0.5271 \pm 0.2884$\\
  90\% & $0.4670 \pm 0.2849$  & $0.4953 \pm 0.2868$   & $0.4747 \pm 0.2870$\\ \hline
  \end{tabular}
  \caption{\label{tab: results_anova_robustness} Summary of mean$\pm$standard deviation of $p$-values of the T-ANOVA, Heo's ANOVA, and PERMANOVA in 100 simulations. Top half: In each simulation, the test is used to compare Group 1, 2, 3 of respective $n_1, n_2, n_3$ random samples are generated with a pre-specified percentage (90\%, 95\%, 100\%) of 100 points from the original key shape 'contaminated' with topological noise in the form of a much smaller hole next to the keyhole with pre-specified (in such case $n_1 = n_2 = n_3 = 4$) or random locations (in such case $n_1 = n_2 = n_3 = 5$, $n_1 = n_2 = n_3 = 20$, or $n_1 = 5, n_2 = 20, n_3 = 100$). Bottom half: In each simulation, the test is used to compare Group 1, 2, 3 of respective $n_1, n_2, n_3$ random samples are generated with a pre-specified percentage (90\%, 95\%, 100\%) of 100 points from the original key shape with a pre-specified (in such case $n_1 = n_2 = n_3 = 4$) or random (in such case $n_1 = n_2 = n_3 = 5$, $n_1 = n_2 = n_3 = 20$, or $n_1 = 5, n_2 = 20, n_3 = 100$) quarter of the keyhole left}
  \end{table}

\subsubsection{Robustness under variation of noise and hole location} We conduct two studies to assess the robustness of the test when the underlying topological structure is 'contaminated' with heterogeneous topological noise and when the hole location shifts as in Section~\ref{sec: sim_trans_robust}.

We first evaluate the robustness of T-ANOVA under heterogeneity of topological noise. In each of 100 simulations, Group 1, 2, 3 of respective $n_1, n_2, n_3$ random samples are generated with a pre-specified percentage (90\%, 95\%, 100\%) of 100 points from the original key shape 'contaminated' with topological noise in the form of a much smaller hole next to the keyhole with pre-specified (in such case $n_1 = n_2 = n_3 = 4$) or random locations (in such case $n_1 = n_2 = n_3 = 5$, $n_1 = n_2 = n_3 = 20$, or $n_1 = 5, n_2 = 20, n_3 = 100$). The 4 possible locations of the topological noise in each group are the same as Figure~\ref{fig: sim_key}. We use the T-ANOVA test to compare PDs across the three groups. We expect the test to stay robust to the topological noise. Table~\ref{tab: results_anova_robustness} (top half) shows the results of the T-ANOVA test in comparison with standard PERMANOVA and the topological ANOVA proposed by \citet{Heo2012}.

We then evaluate the robustness of T-ANOVA under variation of hole location. In each of 100 simulations, Group 1, 2, 3 of respective $n_1, n_2, n_3$ random samples are generated with a pre-specified percentage (90\%, 95\%, 100\%) of 100 points from the original key shape with a pre-specified (in such case $n_1 = n_2 = n_3 = 4$) or random (in such case $n_1 = n_2 = n_3 = 5$, $n_1 = n_2 = n_3 = 20$, or $n_1 = 5, n_2 = 20, n_3 = 100$) quarter of the keyhole left. The 4 possible variants of the key shape in each group are the same as Figure~\ref{fig: sim_key} . We use the T-ANOVA test to compare PDs across the three groups. We expect the test to stay robust to this change in structure, which is not topological in nature. Table~\ref{tab: results_anova_robustness} (bottom half) shows the results of the T-ANOVA test in comparison with the other tests.

\begin{table}[b!]
  \begin{tabular}{ccc}
  \hline
  \multicolumn{3}{c}{Pre-specified Location: $n_1 = n_2 = n_3 = 4$}\\\hline
   Percentage of Points in Key Shape & T-ANOVA & PERMANOVA  \\\hline
  100\% & $113.92 \pm 1.58$ & $8.36 \pm 0.10$\\
  95\% & $114.06 \pm 1.53$  & $8.39 \pm 0.08$\\
  90\% & $114.34 \pm 1.82$  & $8.34 \pm 0.06$\\ \hline
  \multicolumn{3}{c}{Random Location: $n_1 = n_2 = n_3 = 5$} \\\hline
     Percentage of Points in Key Shape & T-ANOVA & PERMANOVA  \\\hline
  100\% & $116.46 \pm 1.39$ & $9.02 \pm 0.07$\\
  95\% & $115.82 \pm 1.57$  & $9.04 \pm 0.13$\\
  90\% & $115.91 \pm 1.45$  & $9.16 \pm 0.26$\\ \hline
  \multicolumn{3}{c}{Random Location: $n_1 = n_2 = n_3 = 20$}\\\hline
     Percentage of Points in Key Shape & T-ANOVA & PERMANOVA \\\hline
  100\% & $127.74 \pm 1.49$ & $36.82 \pm 0.34$\\
  95\% & $128.07 \pm 1.68$  & $36.80 \pm 0.22$\\
  90\% & $128.66 \pm 2.66$  & $36.76 \pm 0.26$\\ \hline
  \multicolumn{3}{c}{Random Location: $n_1 = 5, n_2 = 20, n_3 = 100$} \\\hline
     Percentage of Points in Key Shape & T-ANOVA & PERMANOVA  \\\hline
  100\% & $159.93 \pm 2.59$ & $132.14 \pm 0.58$\\
  95\% & $159.65 \pm 2.74$  & $132.54 \pm 0.82$\\
  90\% & $160.45 \pm 2.77$  & $133.32 \pm 2.56$\\ \hline
  \end{tabular}
  \caption{\label{tab: anova_time} Summary of mean$\pm$standard deviation of computational time in seconds for a million transpositions by the T-ANOVA and PERMANOVA in 100 simulations under the topological noise setting.}
  \end{table}

\subsubsection{Computational time} Table~\ref{tab: anova_time} shows the means and standard deviations of computational times for one million transpositions under the topological noise setting (the hole location and sensitivity studies have similar computational times, so we only present one setting here). Just like the two-sample test, T-ANOVA shows steady growth of computational time as group sample sizes increase, in comparison with the sharp time growth of PERMANOVA. Heo's ANOVA is fast as it runs a univariate ANOVA on the dimensionality-reduced PDs. 

\subsection{Summary} The results show that the performance of our T-ANOVA test is comparable with the two baseline methods in terms of robustness under variation of topological noise and hole location, as well as sensitivity in detecting differential hole presence among multiple groups. In comparison with PERMANOVA, the advantage of the transposition approach of T-ANOVA shows up in the steady growth of computational time as group sample sizes increase. Although T-ANOVA is comparable in performance as Heo's ANOVA, it does not require dimensionality reduction of PDs. More importantly, it has a natural framework for distance-based clustering, which we illustrate in the Application section.
 
\section{Application}
\label{sec: application}

\begin{figure}[b!]
  \includegraphics[width = 1\linewidth]{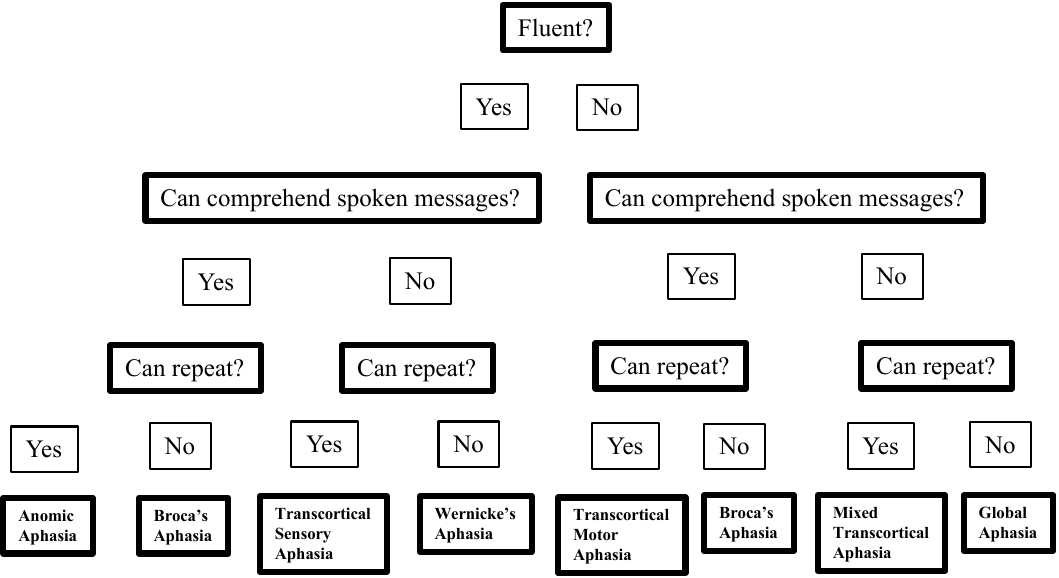}
  \caption{\label{fig: aphasia_subtypes}Traditional aphasia subtypes according to binary categories of fluency, comprehension, and repetition.}
\end{figure}

Stroke is the leading cause of severe adult disability in the United States \citep{Tsao2022}. A left-hemisphere stroke commonly leads to aphasia, a speech-language disorder often classified into subtypes according to behavioral symptoms. Traditional subtypes of aphasia are determined through the Aphasia Quotient (AQ) subtest scores of the Revised Western Aphasia Battery (WAB-R) \citep{Kertesz2007} that assess speech and language abilities such as spontaneous speech fluency, auditory comprehension, repetition, and naming performance. These scores binarize the patients into categories. For instance, the spontaneous speech fluency score ($\ge 5$ vs. $\le 4$) is a rating based on subjective evaluation mostly about quantity and grammaticality of output along with other features, such as word-finding difficulty, paraphasias, and hesitations. It separates individuals into fluent and non-fluent categories. Eight traditional subtypes thus arise from the binarized categories of fluency, comprehension, and repetition (Figure~\ref{fig: aphasia_subtypes}). 
Studies over the years have addressed WAB subtyping issues since its initial version in 1982 and proposed new ways identifying coherent clusters of aphasia subtypes \citep{Ferro1987, Crary1992, John2017, Fromm2022}. 
Unsupervised learning approaches such as K-means clustering has been applied to behavioral scores beyond WAB-R to redefine aphasia subtypes \citep{Fromm2022}. There is, however, a lack of exploration on aphasia subtyping via clustering brain network features. Our goal is aimed at identifying patterns of damage in the brain networks that lead to overlapping behavioral deficits. This study takes a topological angle at the clustering and inference of the resting-state functional brain networks of aphasic individuals, and summarizing basic statistical characteristics of the WAB-R AQ subtest scores of the clusters.

\subsection{Data acquisition and preprocessing}

The rs-fMRI data were acquired from 103 participants with aphasia resulting from a single ischemic or hemorrhagic stroke involving the left hemisphere on a Siemens Prisma 3T scanner with a 20-channel head coil located at the Center for the Study of Aphasia Recovery at the University of South Carolina. The following imaging parameters of images were used: a multiband sequence (x2) with a $216 \times216$ mm field of view, a $90 \times 90$ matrix size, and a 72-degree flip angle, 50 axial slices (2 mm thick with $20 \% $ gap yielding 2.4 mm between slice centers), repetition time TR =1650 ms, TE=35 ms, GRAPPA=2, 44 reference lines, interleaved ascending slice order. During the scanning process, the participants were instructed to stay still with eyes closed. A total of 370 volumes were acquired. 

The preprocessing procedures of the rs-fMRI data include motion correction, brain extraction and time correction using a novel method developed for stroke patients \citep{Yourganov2018}. The Realign and Unwarp procedure in SPM12 with default settings was used for motion correction. Brain extraction was then performed using the SPM12 script pm\_brain\_mask with default settings. Slice time correction was also done using SPM12. The mean fMRI volume for each participant was then aligned to the corresponding T2-weighted image to compute the spatial transformation between the data and the lesion mask. The fMRI data were then spatially smoothed with a Gaussian kernel with FWHM= 6 mm. To eliminate artifacts driven by lesions, a pipeline proposed by \citet{Yourganov2018} was applied on the the rs-fMRI. The FSL MELODIC package was used to decompose the data into independent components (ICs) and to compute the Z-scored spatial maps for the ICs. The spatial maps were thresholded at $p < 0.05$ and compared with the lesion mask for the participant. The Jaccard index, computed as the ratio between the numbers of voxels in the intersection and union, was used to quantify the amount of spatial overlap between the lesion mask and thresholded IC maps, both of which were binary. ICs corresponding to Jaccard index greater than $5\%$ were deemed significantly overlapping with the lesion mask and then regressed out of the fMRI data using the fsl\_regfilt script from the FSL package. By applying the automated anatomical labelling (AAL) atlas, 116 regions of interest (ROIs) were created and used as nodes in the brain networks subsequently constructed. 

The Aphasia Quotient, a score strongly related to the overall lesion damage in brain, was measured in the participants. In terms of behavioral measures, the following WAB-R subscores were used to measure performance of participants in fluency, comprehension, repetition, object naming, and sentence completion: Information Content, Fluency Rating, Spontaneous Speech Rating, Comprehension Yes/No Questions, Comprehension Auditory Words, Comprehension Sequential Commands, Comprehension Subscore, Repetition Subscore, Object Naming, Word Fluency, Sentence Completion, Responsive Speech, and Naming Subscore. 
    
%

\subsection{Resting-state functional brain network and filtration}

We first constructed resting-state functional brain networks from the rs-fMRI described above. The 116 AAL ROIs served as the nodes of the resting-state functional network of each individual and Pearson's correlation between the BOLD signals at two ROIs serve as their edge weight. A Rips filtration was built on the resting-state functional correlation matrix of each individual. The PDs decoding the birth and death times of 1-cycles in the individual Rip filtrations were then smoothed with the HK representation.  

\subsection{Aphasia subtyping via topological clustering of brain networks}
\label{sec: clustering}

Topological clustering has been applied in different angles to studies of resting-state functional brain networks \citep{Stolz2017, Chung2023}. To the best of our knowledge, this is the first study to explore aphasia subtyping through topological clustering of resting-state functional brain networks. Here we take advantage of the HK representation of PDs and extend the T-ANOVA into a topological clustering scheme where clusters were identified with respect to topological centroids calculated as the functional means of the HK estimates of the PDs representing the brain networks of individuals in the study. We compared the statistical characteristics of the topological clusters to baseline clusters obtained through K-means clustering of the WAB-R subscores. We repeated the clustering process 100 times in each instance and checked for consistency across repetitions. Three topological clusters had the overall best fit so we compared the results of three baseline clusters with them. 

\begin{figure}[t!]
  \includegraphics[width = 1\linewidth]{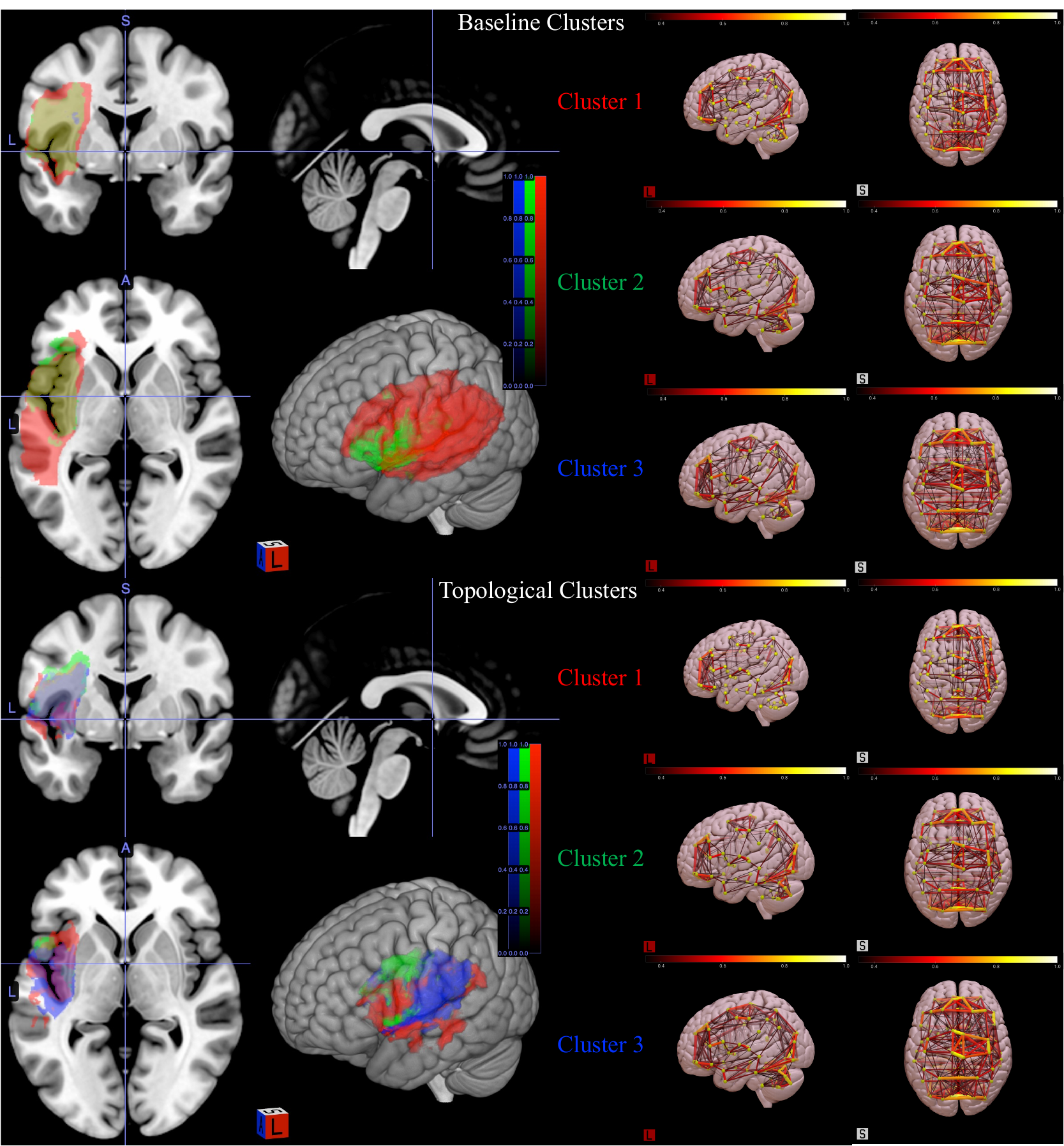}
  \caption{\label{fig: lesion_conn}The lesion map (left two columns) and average absolute connectivity (right two columns) of three topological and baseline clusters.}
\end{figure} 

\begin{figure}[t!]
  \includegraphics[width = 1\linewidth, trim = 2in 0in 2in 0in]{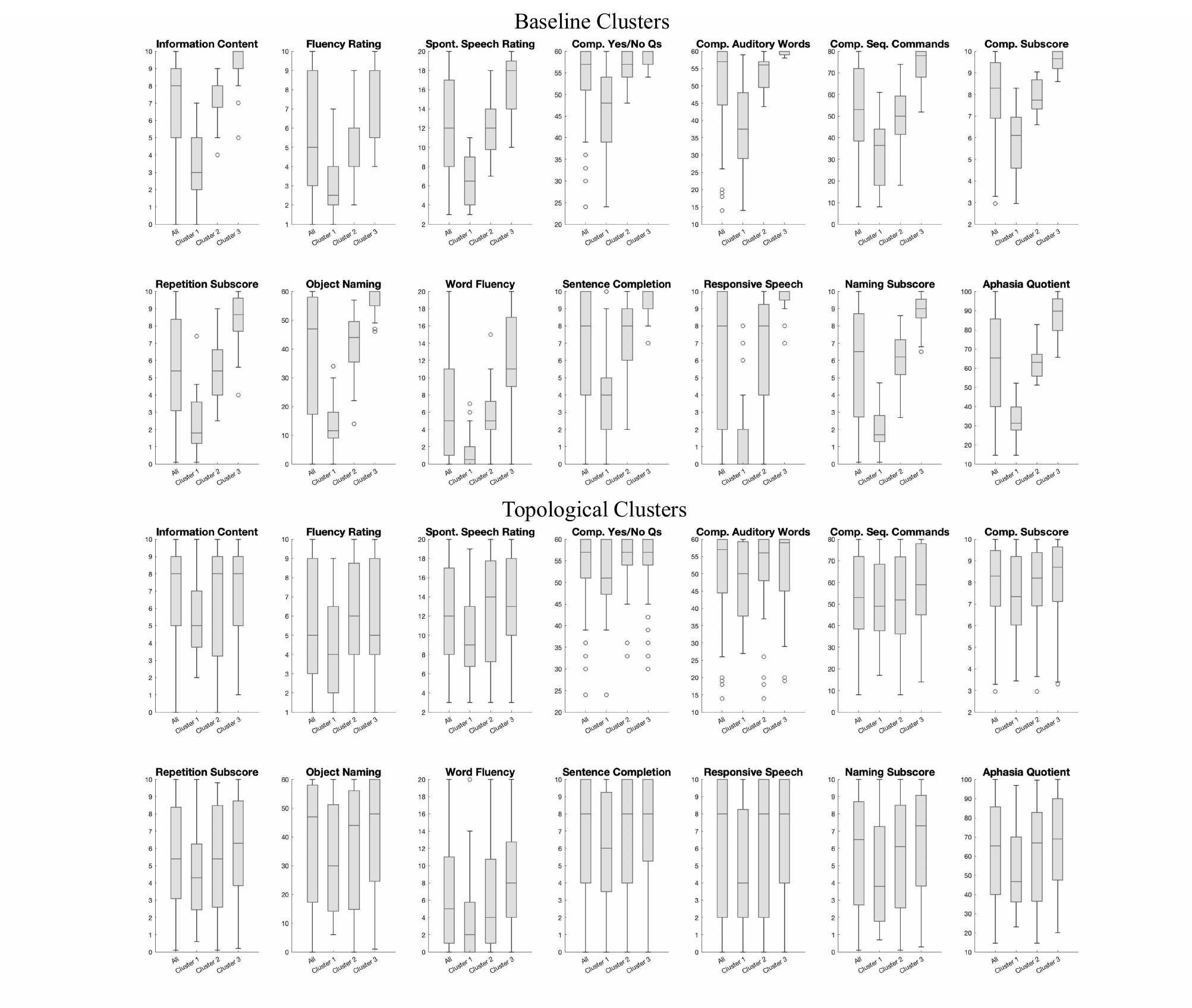}
  \caption{\label{fig: cluster_stats}Box plots of subscores for all participants and those in each of the three clusters.} 
\end{figure}

\begin{figure}[t!]
  \includegraphics[width = 1\linewidth]{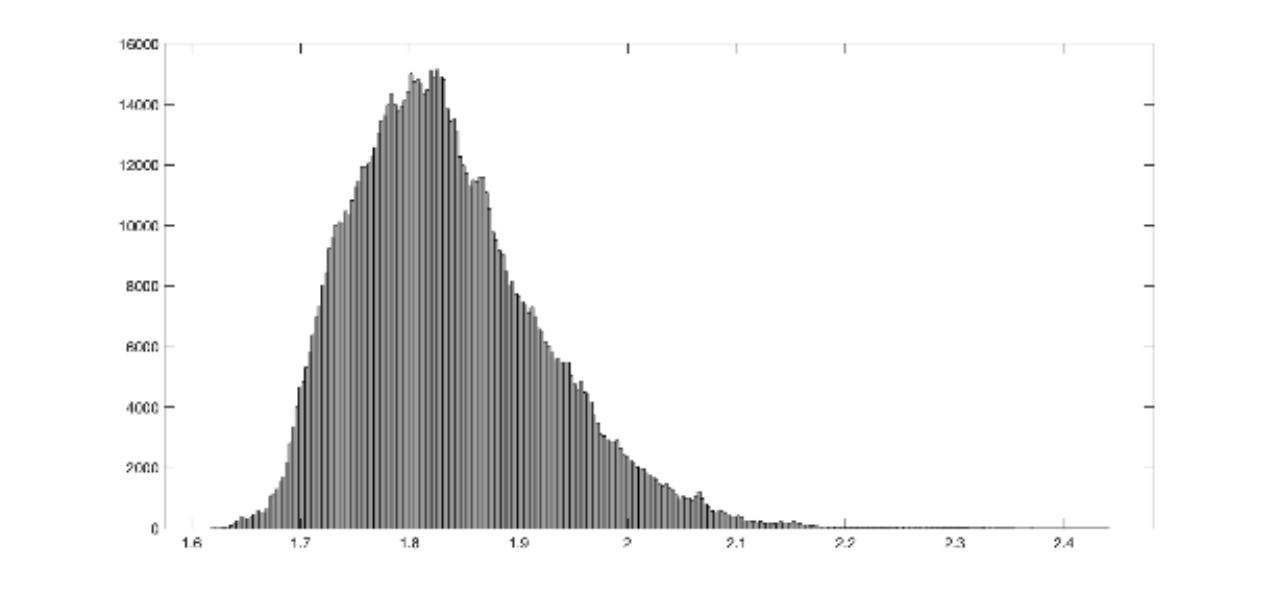}
  \caption{\label{fig: polar_trans}Empirical distribution of the ratio statistic based on $L_2$-distances of HK-estimated 1-dimensional PDs within and between the three clusters over 1 million transpositions.}
\end{figure}

The overall lesion map and average absolute connectivity of three baseline and topological clusters are shown in Figure~\ref{fig: lesion_conn}. The lesion map was created by augmenting stroke lesion damage in the brain of all subjects within each cluster. Note that the three baseline clusters appear to be confounded by the overall lesion extent of the subjects as they show distinctly different lesion extent (Cluster 1 > Cluster 2 > Cluster 3). This is confirmed by the AQ and subscore distributions summarized in Figure~\ref{fig: cluster_stats}, where the AQ score is known to positively correlate with lesion extent and the subscore distributions show a distinct monotone pattern consistent with that of AQ across clusters. On the other hand, the topological clusters do not appear to be confounded by lesion extent as the lesion extent do not vary significantly across the clusters and the subscore distributions do not follow a specific trend with reference to the AQ score. As of the average connectivity, we see different connectivity patterns in the three topological clusters, whereas the baseline clusters show similar connectivity pattens. To confirm that the topological clusters did capture significant statistical difference in brain networks, we also compared the brain networks across different clusters through the T-ANOVA on their HK-estimated PDs. Figure~\ref{fig: polar_trans} shows the empirical distribution of the ratio statistic based on $L_2$-distances of HK-estimated 1-dimensional PDs within and between the three clusters over 1 million transpositions. The observed value of the ratio statistic was 5.4728, yielding a $p$-value of 0 and the conclusion of significant topological difference between the one-dimensional hole presence in the three clusters of brain networks.

\begin{table}[b!] 
\begin{tabular}{c|ccc}
\hline
WAB-R Subscore(s) & Cluster 1 & Cluster 2 & Cluster 3 \\\hline
Comprehension Yes/No Questions & Low & Medium & Medium\\
Comprehension Auditory Words & Low & Medium & High \\
Comprehension Sequential Commands & Low & Medium & High \\
Comprehension Subscore & Low & Medium & High \\
Repetition Subscore & Low & Medium & High \\
Fluency Rating & Low & High & Medium \\
Word Fluency & Low & Medium & High \\
Information Content & Low & Medium & Medium \\
Spontaneous Speech Rating & Low & High & Medium \\
Naming Subscore & Low & Medium & High  \\
Object Naming & Low & Medium & High \\
Sentence Completion & Low & Medium & Medium \\
Responsive Speech & Low & Medium & Medium \\
\hline
\end{tabular}
  \caption{\label{tab: char}Pattern of median and interquartile range of WAB-R subscores across three topological clusters/subtypes.}
\end{table}

Now, using the three topological clusters as a basis for exploring aphasia subtypes, behavioral measures in the form of WAB-R subscores across the three clusters/subtypes have pattern of median and interquartile range summarized in Table~\ref{tab: char}. In terms of the three categories (fluency, comprehension, repetition) used for traditional aphasia subtyping, the comprehension subscores (Comprehension Yes/No Questions, Comprehension Auditory Words, Comprehension Sequential Commands, Comprehension Subscore), Repetition Subscore, and Word Fluency show an overall pattern low-medium-high in medians across Cluster 1, 2 \& 3, with the exception of Comp. Yes/No Qs which sees some leveling off in Cluster 2 \& 3, whereas Fluency Rating shows a low-high-medium pattern across the three clusters. The low, medium, and high are all in comparison to the median over all subjects.

\section{Discussion}

In this study, we established a topological inference framework based on HK representation of PDs. Although it does not require the PDs to be extracted from a specific type of data, we centered the application of the methods around group comparison of PDs from brain networks. But simulating brain networks with holes is not straightforward, so we used point clouds from images with an underlying shape in illustration and simulation studies. We also extended the framework to topological clustering of brain networks with application to subtyping individuals with post-stroke aphasia. 

Methodologically, the topological inference framework filled a few gaps left from our previous works. As we pointed out in Section~\ref{sec: methods}, the spectral transposition test generalizes the permutation test proposed by \citet{Wang2018} that compares single-trial signals by permuting coefficients respective of Fourier basis functions. We can now permute the Fourier coefficients of the HK estimates of two groups of PDs, which may come from multi-trial univariate or multivariate signals. Thus the framework is now not only applicable for single-trial univariate signals, but also for multi-trial univariate and multivariate signals. Furthermore, we now have T-ANOVA that can compare the topological features of multi-group univariate and multivariate signals without further reducing the dimensionality of the features. The multi-group transposition approach can also be used for speeding up ANOVA procedures in non-topological settings. Resampling is also of high relevance to deep learning. Since the power of deep learning is constrained in small sample schemes, data augmentation methods are needed to increase the training data by resampling \citep{Huang2021}. In future studies, the proposed spectral permutation method can be easily adapted for deep learning where the input is augmented persistence features reconstructed from resampled HK coefficients of PDs. 

Since the analytical paradigm proposed in the methods section was already complicated, we featured topological clustering only in application. In the application of topological clustering and inference to subtyping individuals with post-stroke aphasia, one would argue that three clusters may be too sparse for actual clinical interpretation even though three clusters were empirically determined to have the best fit. Future studies can refine the approach by exploring more clusters, e.g. matching the number of traditional subtypes to see if they have any consistency. 

\section*{Acknowledgments} The authors would like to thank Dr. Roger Newman-Norlund for facilitating access to the fMRI dataset used in this study and Dr. Moo Chung for helpful discussions on the early versions of the manuscript. Funding sources: NIH R01DC017162 and R01DC01716202S1 (PI: RHD). Author contributions: conceptualization (YW, RHD), statistical analysis (YW, JY), interpretation of results (YW, RHD), writing and editing (All). Compliance with ethical standards: The neuroimaging scans were approved by the Institutional Review Board (IRB) at the University of South Carolina. Conflict of interest and disclosure: None.

\bibliographystyle{apacite}
\bibliography{bib_top_inf_arxiv} 

\end{document}